# Review of Spin–Orbit Coupled Semimetal SrIrO$_3$ in thin film form

Lunyong Zhang[1,3*], Bin Pang[1], Y.B Chen[2*], Yanfeng Chen[1]

1. National Laboratory of Solid State Microstructures and Department of Materials Science and Engineering, Nanjing University, Nanjing 210093, People's Republic of China
2. National Laboratory of Solid State Microstructures and Department of Physics, Nanjing University, Nanjing 210093, People's Republic of China
3. Max Planck POSTECH Center for Complex Phase Materials, Max Planck POSTECH/Korea Research Initiative (MPK), Gyeongbuk 376-73, Korea

Corresponding author: Lunyong Zhang, email: allen.zhang.ly@gmail.com

Y.B Chen, email: ybchen@nju.edu.cn

**Abstract:** Spin–orbit coupling, locking the momentum of an electron to its spin, has been shown essential for giving rise to many novel physical behaviors. SrIrO$_3$ is a typical metallic member of the strong spin–orbit coupling iridate family. Its orthorhombic phase has been confirmed as a paramagnetic semimetal resulted from the interplay among spin–orbit coupling, electron correlation, and crystal field, and was theoretically predicted to host versatile topological phases. This article reviews the current knowledge on the preparation and the tunable properties of orthorhombic SrIrO$_3$ films. Experiments have demonstrated that orthorhombic SrIrO$_3$ can be successfully synthesized as films under substrate lattice constraint without high pressure, and the films frequently display metal-insulator transition due to disorder and weak-antilocalization owing to spin-orbit coupling. The properties of orthorhombic SrIrO$_3$ film are sensitive to the rotation and tilting of the IrO$_6$ octahedral, and consequently can be significantly tuned through strain engineering. Simultaneously, thickness dependent size effect is also remarkable in SrIrO$_3$ films. The accumulated research on SrIrO$_3$ films suggests an urgent demanding for research on superlattices constructed with orthorhombic SrIrO$_3$, to better understand the mechanism of the electron structure evolution, and thus the relevant magnetic states and topological phases in orthorhombic SrIrO$_3$ and its family.

# 1. Introduction

Emergent materials with strong spin orbit coupling (SOC) have excited hot research waves in the condensed matter physics and materials communities, due to the observed or predicted versatile novel physics such as topological states, superconductivity, and quantum critical states. The novel physics originate principally from the competition between the strong SOC ($\zeta_{SO}$) and other basic interactions including the crystal field ($\triangle$) and the electron correlation ($U$), due to their comparable energy scale in these materials. This competition is not founded in the traditionally investigated 3d transition metal compounds where the SOC is often negligible in most physical effects, but formed in the 5d-transition metal compounds typically for example the iridates. As a prototype SOC system, iridates have $\zeta_{SO}$ about 0.4 eV. It is five to ten times over that of 3d elements, compared with the oxide octahedral crystal field ($\triangle \sim 2$ eV) and the electron correlation ($U \sim 0.5$ eV). The strong SOC occupies an essential position in the formation of the proposed and the observed superconductivity [1], Wyle semimetal [2], quantum criticality [3], and quantum Hall effect [4-6] in diverse iridates. Iridates have been deemed as an excellent playground for investigating the exotic physics generated by nonnegligible SOC. For example, the SOC driven $J_{eff} = 1/2$ band structure model (Fig.1b) initially proposed in the two dimensional member of Ruddlesden-Popper (RP) structure type $Sr_{n+1}Ir_nO_{3n+1}$, $Sr_2IrO_4$ (n=1). It has been taken as the beginning for discussing most of the discovered novel physics in perovskite iridates, where strong SOC splits the $t_{2g}$ band into a $J_{eff} = 1/2$ doublet band and a $J_{eff} = 3/2$ quartet band on the basis of the octahedral crystal-field splitting. Then the moderate electron correlation $U$ can further lift the $J_{eff} = 1/2$ doublet band into an upper Hubbard band (UHB) and a lower Hubbard band (LHB), constructing a half filled $J_{eff} = 1/2$ band model [7]. This particular band structure is modulated by the magnitude of Hubbard $U$ and the electron hopping integral, therefore, by the delicate tilting and rotation of $IrO_6$ octahedra from the viewpoint of crystal structure.

Recent years, the perovskite $SrIrO_3$ (n=∞, hereafter abbreviated as $SrIrO_3$) with orthorhombic structure has attracted increasing interests. Owing to the enhanced interlayer hopping induced by the three-dimensionality, in comparison with $Sr_2IrO_4$ and $Sr_3Ir_2O_7$ (n=2), the $J_{eff} = 1/2$ band mixes the $J_{eff} = 3/2$ band at a certain extent in $SrIrO_3$. The $SrIrO_3$ becomes, consequently, a particular paramagnetic semimetal [8, 9]. In contrast, both of the $Sr_2IrO_4$ and the $Sr_3Ir_2O_7$ are insulator with canted antiferromagnetic ordering [9-11]. In physics, the $SrIrO_3$ is, therefore, close to a multi-phase boundary with both the metal–insulator transition and the magnetic background state transition, which are controlled by the competition between the multiple interactions. The relevant investigations are helpful in understanding the evolutionary mechanism of the electronic and the magnetic properties, the modulation of the $J_{eff}=1/2$ band model, the nature of the competition amongst the multiple interactions in iridates and other strong SOC systems. Experiments have observed the weak antilocalization effect[12], validating the strong SOC feature of $SrIrO_3$. Nonlinear Hall effect[13] and angle-resolved photoemission spectroscopy (ARPES) band structure characterizations[14, 15] have proved the semimetallic proposition of $SrIrO_3$ made by the first principle calculations[16]. This semimetallic background state is highly sensitive to structure modulation achieved by for example the substrate strain, as well as sensitive to the quantum size effect[15, 17, 18]. A canted antiferromagnetic ordering can be induced with decreasing the film thickness[19]. These points will be discussed in detail in sections 4, 5 and 6 of this review, respectively.

In addition, the $SrIrO_3$ was proposed to host the Dirac cone in the band structure and can be modified into varied topological phases[16, 20-22] dependent on the broken structure symmetry, which is clearly important for opening up a new research field of topological phase materials based on oxides. At the materials side, the broken structure symmetry is realized by constructing heterostructure based on $SrIrO_3$ and other $ABO_3$ perovskite oxides along particular crystallographic orientations, for example, the [111] direction and the [001] direction. The component layers require thickness at unit cell scale and lattice quality at atomic level, which sets a big challenge to prepare the theoretically supposed samples. Consequently, the topological phase based on $SrIrO_3$ is still an open question in experiment. More investigations on the sample preparation and related topological state characterizations are requisite in future. The present review will discuss that in detail in section 7.

In technologies, the $SrIrO_3$ is a prospective candidate for the oxide spintronics devices applications where the strong SOC for modulating the electron spin degree and the proper carrier conduction are required. Obviously, they are filled by the $SrIrO_3$. Besides, the perovskite structure of the $SrIrO_3$ makes it well match with most of the perovskite transition metal oxides, giving the opportunity to fabricate high-quality oxide devices. The possible topological phases realized in $SrIrO_3$ based materials or heterostructure also make a big potential of achieving the low energy dissipation oxide spintronic devices.

However, bulk SrIrO$_3$ has a BaTiO$_3$-type hexagonal structure at ambient conditions, the perovskite SrIrO$_3$ phase is synthesized only at high pressure over than 4GPa [23]. This problem, fortunately, can be solved through the substrate lattice constraint, so conveniently obtains stable perovskite SrIrO$_3$ film at ambient pressure [24]. The growth techniques of the SrIrO$_3$ film will be discussed in detail in section 3 in the present review. This paper thoroughly reviews the results and understandings obtained so far for the perovskite SrIrO$_3$ films from the growth techniques to physical properties, and simultaneously make the prospective discussions on the topics worth investigated in future. It demonstrates a comprehensive picture for the research community to advance the next explorations of the SrIrO$_3$ as well as the related materials. We noted, during the peer-review process of the present article, a closely recent article written by A. Biswas et al[25] also briefly reviewed some points in the present review. The readers consequently can survey the picture of SrIrO$_3$ from a different view angle.

## 2. Bulk Properties of Perovskite SrIrO$_3$

Naturally, the main motivation for researching a material at low-dimensional scales is to see how the external stimulus, for example, the size confinement, the substrate strain, and the proximity effect modulate the properties of the material. The research, therefore, should begin with an understanding of the bulk properties of the material. Perovskite SrIrO$_3$ is a SrRuO$_3$-type orthorhombic crystal structure with *Pbnm* symmetry, as seen in Fig.1a. The lattice constants are $a$ = 5.56 Å, $b$ = 5.59 Å, and $c$ = 7.88 Å. All of the nearest-neighbor IrO$_6$ octahedra tilt in the opposite sense about the [110] axis near 12°, and moreover, they rotate around the c-axis at about 11° in the same direction if they are in different layers, but in the opposite direction if they are in the same layers, resulting in a unit cell with four formula units. The tilting and rotation of the octahedra simultaneously cause unit cell distortion with two types of Ir–O–Ir angles, at 156.49(8) ° and 153.7(5) °, respectively[23, 26]. In comparison to Sr$_3$Ir$_2$O$_7$ and the Sr$_2$IrO$_4$, the octahedral layers are connected to each other through a sharing of the oxygen atom in the top corner, accordingly constructing a three-dimensional structure for every Ir atom surrounded by six adjacent Ir atoms.

As for the electronic band structure of perovskite SrIrO$_3$, the local-density approximation plus the Hubbard $U$ (LDA+$U$) calculation demonstrates that the bands around the Fermi level are entangled. Adding proper SOC can lift them into the J$_{eff}$ =3/2 band and the J$_{eff}$ =1/2 band [16], the latter is further split into UHB and LHB under the electron correlation interaction $U$, as depicted in the band formation picture in Fig. 1b [7]. By increasing the dimensionality from two in $Sr_2IrO_4$ to three in $SrIrO_3$, electronic hopping increases, the $J_{eff}=3/2$ band and the LHB as well as the UHB are broadened and partially overlapped (Fig.1c), forming a semimetallic state[8, 27]. Figure 1d presents the LDA+SOC+$U$ method for the calculated band structure[16] and Fig.1e presents the band structure obtained through the LDA plus dynamic mean field theory (DMFT) method[28], and they are basically similar. The conductive band is mainly formed by the UHB and the valence band is mainly formed by the LHB. There is a small amount of mixing between them around the Fermi level. In addition, the LHB is also mixed by the $J_{eff}=3/2$ band [14, 28]. This proposed band structure evolution is in agreement with the optical conductivity spectra (OCS) obtained from the $SrIrO_3$ film on MgO substrate, $Sr_2IrO_4$ and $Sr_3Ir_2O_7$ single crystals (Fig.6a) [8], but bears some discrepancies with the OCS results collected from $[(SrIrO_3)_m(SrTiO_3)]$ superlattice (Fig.6b-6g)[29] and with the APRES probed band width of $SrIrO_3$ films[14, 15] (Fig.7a and 7b). This will be discussed in detail in section 4.

Clearly, both of the LDA + SOC + U (Fig.1d) and the LDA+ DMFT (Fig.1e) methods predicted a non-trivial Dirac cone-like node. It actually was also proposed in the calculation using the generalized gradient approximation plus $U$ method (GGA+$U$, shown as the dot line in Fig.1e) and the tight-binding approximation method which considers only the nearest and next-nearest neighbor Ir–Ir direct hopping between $J_{eff}$ = 1/2 orbitals [16]. The node constructs the basis for predicting how the $SrIrO_3$ can be modified into various topological states[16, 20, 21]. It was theoretically demonstrated that the symmetry protection of the Dirac node could be broken by different hetero-epitaxial structures due to its origination from the reflection symmetry operation of the crystal structure[20, 30] (A most recent work reported the Dirac node is protected by the n-glide symmetry not the mirror symmetry[30]). In experiment, a nearly linear electron band of about 50 meV below the Fermi level has been observed on the film on a (001)-LSAT substrate[14], however the Dirac node has not been confirmed beyond doubt[14, 15] (see section 4 for detail).

Figure 1b and 1c also demonstrate that the Fermi level crosses not only the conductance band above the linear node but also the top of the valence bands. This semimetal picture was recently verified by ARPES measurements (Fig.7) [14, 15, 31] and the nonlinear Hall effect in an electrical transport measurement of $SrIrO_3$ films[13]. Since both the conductance band and the valence band are slightly crossed by the Fermi level, forming the so-called electron pocket and hole-like pocket, the semimetallic ground state can be tuned through chemical doping and lattice strain. The latter stimulation is the basis of most existing investigations of SrIrO$_3$ films.

Because high pressure is necessary for synthesizing the perovskite SrIrO$_3$ in bulk state, only a few experimental works were carried out on bulk SrIrO$_3$[23, 26, 32, 33]. Consistent with the band structure prediction of a semimetallic ground state, bulk perovskite SrIrO$_3$ resistivity is decreased with temperature, and demonstrates a semimetallic feature of weakly temperature-dependence in the measurement region from ~2 K to 300 K[32, 33], as revealed in Fig.2a. At high temperature regime, the resistivity is roughly linearly dependent on temperature (Fig.2a). It is evolved as $T^2$ dependence at low temperature regime (Fig.2b), suggesting a typical Fermi liquid metal behavior. Magnetoresistance (MR) effect is persistent in bulk samples at low temperature, as seen in Fig.2a. It should be noted that the metal-insulator transition (MIT) and negative MR were often observed in SrIrO$_3$ samples in both the bulk and the film states, but it is not the case for the polycrystalline samples of Fig.2. Therefore, we can understand the MIT is an extrinsic behavior of SrIrO$_3$. The negative MR right suggests the carrier localization mechanism of the MIT, due to disorder (see section 5.2 for detail). The MR is parabolically dependent on the magnetic field in a simply metal. Fig. 2c indicates a MR smaller than the fit MR according to the $H^2$ law at low fields. At high field regime, the MR is approximately linearly increased with $H$, in both the $B//I$ geometry and the $B\perp I$ geometry (Fig.2d and 2e). This is perhaps a result of the quantum linear positive MR model proposed by A. A. Abrikosov[34] for a linear momentum gapless dispersion band structure, which might correspond to the nearly linear (Dirac) node shown in Fig.1b. If so, unsaturated linear positive MR at the even higher field would be expected.

Fig2.f gives out the Hall coefficient ($R_H$) of the polycrystalline SrIrO$_3$ sample for Fig.2d and 2e. $R_H$ is negative and decreased with temperature until about 25K, and then keeps roughly constant below 25K. Electron carrier therefore dominates the charge transport behavior. In fact, the non-linear Hall effect consistent with the semimetallic feature of SrIrO$_3$ was observed in the films grown on (001)-STO substrates, where the $R_H$ is consistently negative[13]. The temperature dependence of $R_H$ reveals that the electron can be easily thermal excited due to the small splitting between the LHB band and the UHB bands. This is partially shown as well in the optical conductivity spectra (Fig.2g), where the α peak (LHB→UHB excitation, Fig.1c) and the β peak (J$_{eff}$ 3/2→UHB excitation) are well developed, and the peaks are wider at high temperature. The well-developed α peak indicates that the LHB and the UHB are not fully overlapped at least, as predicted by the theoretical model. A metallic Drude like tail were observed as well, consistent with the direct current conductivity of $SrIrO_3$.

Bulk $SrIrO_3$ is paramagnetic, as shown in the Fig.2d that no transition was observed in the magnetic susceptibility curve and the magnetization was linearly dependent on the magnetic field for both of the polycrystalline samples without MIT gained by Blanchard et.al[32] and by Fujioka [33]. This paramagnetic feature was also reproduced in the polycrystalline sample with MIT[26]. A modified Curie–Weiss law, $\chi = \chi_0 + C/(T-\vartheta) + DT^2$, where $\chi_0$, $C$, $T$, and $\vartheta$ are temperature-independent magnetic susceptibility constant, Curie constant, temperature and paramagnetic Curie temperature, respectively. The $T^2$ term is assigned to the higher-order temperature-dependent term of the Pauli paramagnetism[32]. The effective moment of each Ir atom was accordingly derived as 0.062 $\mu_B$ for the Blanchard's sample but 0.156$\mu_B$ for the Fujioka's sample. The difference between the two samples' effective moments perhaps comes from the disorders. Nevertheless, they are consistent with the reported effective moments of the $Sr_2IrO_4$ [35] and the $Sr_3Ir_2O_7$ [36] at magnitude order, which could be deemed as a hint of the validity of the $J_{eff}=1/2$ model in the $SrIrO_3$. On the other hand, it is distinctly large smaller than that for a local S=1/2 electron, and much smaller than the idea value of an unpaired $5d$ electron in the $Ir^{4+}$ cation according to the Hund's rule, ~1.73$\mu_B$. This is attributed to the strong SOC and the strong hybridization between Ir-$5d$ and O-$2p$ orbits, as proposed in many other iridates[36, 37].

## 3. Thin Film Growth and Microstructure

In literatures, $SrIrO_3$ film can be prepared through varied methods including the pulsed laser deposition (PLD), the molecular beam epitaxy (MBE), the sputtering technique, and the metalorganic chemical vapor deposition (MOCVD). The pulsed laser deposition (PLD) technique is undoubtedly the most adopted method recently for epitaxial growing of perovskite $SrIrO_3$ films, for example see Refs.[ [12], [17], [18], [38-40]]. With PLD, high-quality perovskite $SrIrO_3$ films with a two-dimensional growth mode can be conveniently prepared on substrates such as $SrTiO_3$ (STO), $LaAlO_3$ (LAO), $DyScO_3$ (DSO), $GdScO_3$ (GSO), and $NbScO_3$ (NSO) (for example, see the atomic force microscopy (AFM) image in Fig.4a where large-width step flows are distinctly visible), by using well surface-treated substrates and $O_2$ gas background atmosphere. The growth conditions need to be properly controlled, for example, setting the target-substrate distance at about 70mm, laser frequency often as 2Hz. The substrate temperature is typically around 700 °C [12,18], although it is also applicable at lower temperatures to ~500 °C [38] and higher to ~800 °C from the viewpoint of growing flat surface film. However, it has been reported that substrate temperatures higher than 650 °C may cause serious Ir deficiency in the film, increasing the resistivity, and even transforming the film into an insulator[38]. That is caused by the phase transformation from the $SrIrO_3$ to the $Sr_2IrO_4$ due to the decreased stability of $SrIrO_3$ with substrate temperature increasing[41]. The corresponding temperature for transformation is dependent on the specific conditions for sample making.

For the MBE method[14,15], distilled ozone is used as the oxidant with a background pressure about $10^{-6}$ torr and the substrate temperature around 500 °C has been reported[15]. Compared with the PLD, a typical feature of the MBE method is the lower growth speed. Recently, Li et al. reported to epitaxially grow the $SrIrO_3$ film on (001)-$SrTiO_3$ substrate with sputter techniques[42]. They adopted the radio frequency (RF) magnetron sputtering with the on-axis geometry, with the substrate temperature as 610 °C and the target-substrate distance as 40mm. A mixture of Ar and $O_2$ with ratio of 1 to 2 and with total pressure at 210 mTorr was used as the background atmosphere. Interesting is that they used a sputtering target of $Sr_4IrO_6$ composition, indicating the non-equilibrium feature of the growth. In the MOCVD method [43,44], the substrate temperature was 600°C to 700°C; the source materials were $Sr(C_{11}H_{19}O_2)_2(C_8H_{23}N_5)_x$, $Ir(C_7H_9)(C_6H_8)$, and $O_2$ gas. By controlling the concentration of the input source gas, a deposition rate of 50–70 nm/h could be realized under reactor pressure $1.3 \times 10^3$ Pa and a total gas flow rate of 600 $cm^3$/min.

Perovskite $SrIrO_3$ has pseudocubic represented lattice constants $a_c=b_c \approx \frac{1}{2}\sqrt{a^2+b^2}$=3.942 Å and $c_c = \frac{1}{2}c$=3.94 Å (see Fig.5f), and good lattice matching with most of the common pseudocubic structure substrates, leading to the full lattice strain constraint in a wide range of film thicknesses, which supply the possibility to investigate the effect of strain on $SrIrO_3$ film properties. Varied routes have confirmed the full lattice strain state in $SrIrO_3$ film. The coherent interface was confirmed by high-resolution TEM [12,45] (for an example, see Fig.3c). Figure 4a gives the X-ray reciprocal space mapping (RSM) images of $SrIrO_3$ films grown on the (001)-NSO [18], (001)-STO[13], and (110)-$NdGaO_3$ (NGO) [45] substrates which introduce a lattice strain from 1.2% to -2.4%. Obviously, the diffraction spots of $SrIrO_3$ are located at the same in-plane $Q_x$ or $Q(110)$ as those of the substrate spots. TEM selected-area electron diffraction patterns of the films on the (001)-STO and the (001)-LSAT substrates only demonstrates split diffraction spots with high indexes along the out plane direction, as shown in Fig.4b and 4c, as well as giving out the micro-scale evidence of the full lattice constraint.

In principle, the orthorhombic phase of a SrIrO$_3$ film can be assigned through XRD. However it is dangerous in practice through only XRD because we noted that the XRD patterns of the hexagonal SrIrO$_3$ films and the orthorhombic SrIrO$_3$ film are similar and cannot be easily distinguished in some cases. At the beginning of the investigations of SrIrO$_3$ films, the X-ray pole figure was adopted to confirm the orthorhombic phase. It was shown that the pole figures of SrIrO$_3$ film grown on (001)-SrTiO$_3$ by the MOCVD method[44] and on the MgO substrate grown by the PLD method[8] are four-fold symmetric, suggesting the orthorhombic phase. In contrast, the pole-figure is six-fold symmetric for the film grown on (111)-SrTiO$_3$ substrate by the MOCVD method, corresponding to the hexagonal phase[44]. Zhang et al. has investigated this point through TEM method and demonstrated the direct evidence of that orthorhombic SrIrO$_3$ can be synthesized as film without high pressure. The SrIrO$_3$ films grown on (001)-STO substrates and (001)-LSAT substrates manifest selected-area electron diffraction (SAED) patterns[24] similar to those of the orthorhombic SrRuO$_3$ film[46], confirming the orthorhombic phase of the film, as seen in Fig.5a to 5c [24].

The lattice strain in epitaxial film would gradually relax after a critical thickness through the generation of element vacancy, dislocation and even grain boundary[24, 39]. Chung et al. first noted in SrIrO$_3$ film that a high-quality phase and a relatively poor-quality phase were coexist when the film was thicker than a certain thickness[39]. It should be kept in mind that the orthorhombic phase is only stable in a thin layer contacting with the substrate for a thick SrIrO$_3$ film. As shown in Fig.3e, stripe-like Moire fringes were observed in the second layer, which is induced by the partial overlap of two neighboring grains. The relaxed phase in the second layer is the hexagonal SrIrO$_3$ stable at room pressure[24]. The critical thickness would decrease with increasing the substrate strain, up to around 40nm on the SrTiO$_3$, NdGaO$_3$, GdScO$_3$ and DyScO$_3$ substrates (Fig.3d)[45], and at least 20 nm on the LAO substrate due to the larger lattice mismatch (Fig.3e)[24]. Consequently, the film must be thin enough to keep the lattice perfection and the desired orthorhombic phase. In most cases, a thickness below 10 nm is generally recommended. Moreover, bear SrIrO$_3$ thin films would suffer degradation in air and are highly prone to damage in the lithographic process, a SrTiO$_3$ cap layer has been suggested to overcome the problem[40].

It is possible to form two types of domains labeled as A and B (shown in Fig.5d) in orthorhombic perovskite $SrIrO_3$ films, which can be inspected from the additional weak diffraction spots shown in Fig.5a, 5b, 5d, and 5h. Figure 5d and 5h also indicate that the two domains are rotated 90° toward each other in the substrate plane, and this forms the rotational boundaries illustrated in the TEM contrast images (Fig.5g). Their orientation matches the relationship with the substrate as A: [001](110) $SrIrO_3$ // [100](001) $SrTiO_3$ and B: [001](110) $SrIrO_3$ // [010](001) $SrTiO_3$, schematically shown in Fig.5e[24], which is in the same model with the $SrRuO_3$ film on a (001)-STO substrate[47]. The mechanism for the growth in the [110] direction instead of the [001] direction could be ascribed to the lattice constant $c_c$ which matches the STO lattice constant ($a=b=c=3.904$ Å) more than the $a_c$ and $b_c$ in orthorhombic $SrIrO_3$ along the pseudocubic [001]. This double-domain feature was observed in the $SrIrO_3$ film on the (001)-LSAT substrate as well[14], and should be carefully considered in the investigations of orientation dependent properties such as anisotropic transport measurement and ARPES band structure characterization.

High-quality perovskite $SrIrO_3$ film can be generally grown with a two-dimensional or layer-by-layer growth mode on good lattice-matched substrates (Fig 3a), when the substrate surface is well cleaned and stepped. From a practical viewpoint, annealing at about 1000°C is facile to prepare the stepped surface for the substrates STO [48], DSO [49], GSO, and NSO, but difficult for the LSAT substrates.

## 4. Band structure of thin film

As demonstrated in Section 2, the band structure calculations predicted a non-trivial Dirac cone-like node crossed by the Fermi level as an electron pocket, accompanied by hole pockets, in the bulk perovskite $SrIrO_3$ [16, 28]. On the other hand, due to the size effect and the lattice modulation by substrate strain, the band structure of a film might be varied from that of bulk $SrIrO_3$.

Moon et al. investigated the band structure of $SrIrO_3$ film grown on MgO substrate through optical conductivity spectra (OCS) measurements (Fig.6a), in comparison with the spectra of $Sr_2IrO_4$ single crystal and $Sr_3Ir_2O_7$ single crystal[8, 50]. The electron excitation peaks $\alpha$ (LHB→UHB) and $\beta$ ($J_{eff}=3/2$→UHB) are distinctly observed for the latter two samples, and the peaks shift towards low energy with increasing the dimensionality of the samples. Simultaneously, the width of the peaks was increased. For $SrIrO_3$, the $\alpha$ peak becomes a Drude like response in the energy region below 0.5eV due to the mixing of LHB and UHB. Moreover, the tail of the Drude-like response were enhanced with temperature decreasing, on the contrary with the behavior of conventional metals, indicating the band structure of $SrIrO_3$ is a result of the complex competition amongst multiple interactions[50]. S.Y. Kim et al. alternately investigated the dimensionality dependent band structure evolution of RP iridates through artificially controlling the layer numbers $m$ in the $[(SrIrO_3)_m(SrTiO_3)]$ superlattice[29]. It was discovered that the OCS $\sigma_1(\omega)$ of the samples bears similarity with those of $Sr_2IrO_4$ and $Sr_3Ir_2O_7$, with $m$ as 1 and 2, respectively (Fig.6b-6c). The $SrIrO_3$ film sample with $m=\infty$ exhibits well developed $\alpha$ peak and $\beta$ peak, similar to the polycrystalline sample's OCS shown in Fig.2g. Increasing $m$ shifts the $\alpha$ peak and the $\beta$ peak to lower energies (Fig.6e) and increase the low energy spectra weight (Fig.6f), consistent with the increase in direct current conductivity and the calculations made by B.J. Kim [51], corresponding to the reduction of $U$ and increased mixing of the bands UHB/LHB and LHB/$J_{eff}$=3/2. However, the peaks become shaper with $m$ increase (Fig.6g), a sharper $\alpha$ peak was manifested in $SrIrO_3$ not in $Sr_2IrO_4$, that implies decreased bandwidth, does not agree with the results collected by Moon and with the band structure model discussed in section 2. The calculations made by B.J. Kim gives out a converse result as well. On the other hand, S.Y. Kim's results are in coincidence with the results of ARPES band structure characterizations shown as following.

Figure 7a-7g summarized the featured patterns of the ARPES probed band structure of $SrIrO_3$ films grown on the (001)-LSAT substrate by Nie et al[14], and on the (001)-$SrTiO_3$ substrate by Liu et al. [15]. Generally, the $J_{eff}$=1/2 and $J_{eff}$=3/2 bands and their dispersions, in company with their partial overlap, were validated, as seen in both the Fig.7a and Fig.7d, independent on the specific substrate. Split hole-like pockets were revealed at the center Z (Fig.7a and 7b) and the corner R (Fig.7d and 7f) of the Brillouin zone, and electron pockets were probed at the U point in the Brillouin zone (Fig.7c and 7e)[14, 15]. They construct a semimetallic state. As shown by the Fermi surface patterns in Fig.7g and 7h (Fig.7h is obtained by Yamasaki et al. from $SrIrO_3$ film grown on Nb-doped (001)-STO substrate[31]), where the electron pockets and hole pockets contribution are separated. The electron pockets display nearly linear but anisotropic dispersions, which generate carriers with high Fermi velocities, in comparison with the heavy effective masses of the hole-like pockets quasiparticles[14, 15, 31] (Table 1). This explains why electrons, not holes, dominate the transport of $SrIrO_3$, as seen in the Hall effect measurements[13, 14, 18], despite the expected high

density of state for the heavy hole pockets at the Fermi level. These features were reproduced by the first principle calculations with SOC (for example see Fig.7g), but it was reported that the Coulomb correlation energy $U$ is not necessary or has weak effect on the calculated band structure. That was deemed as an evidence of the critical role of the SOC in the SrIrO$_3$[14,15].

Quite importantly, a nearly linear dispersed electron pocket at the U point is a Dirac cone in theory. It was seemly observed by Nie et.al on the film on (001)-LSAT substrate[14] from the data shown in Fig.7c. However, Liu et.al argued that the band is gaped there on the film on (001)-SrTiO$_3$ substrate based on the 2$^{nd}$ derivative of the original ARPES data (Fig.7b) [15]. Actually, both of them probed similar band pictures. Both of Fig.7c and Fig.7e give out diffused strength cone at the U point with energy lower than the Fermi level about 0.05ev, but both of their 2$^{nd}$ derivative band structures (Fig.7a and Fig.7b) did not show a distinctive non-gapped cone there. However, these results are still limited by their energy resolution; they should not be deemed as the definite conclusion about the existence of the Dirac cone. ARPES characterizations with high-energy resolution are still necessary. In particular, most recent structure refinements revealed the breaking of the n-glide symmetry in the strained SrIrO$_3$ films, which removes the Dirac node[30].

In addition, the J$_{eff}$ =1/2 and J$_{eff}$ = 3/2 bands were shown to be mixed to a different extent in the various crystal orientations [14,15], suggesting weakened SOC effect in comparison with those in the Sr$_2$IrO$_4$ and Sr$_3$Ir$_2$O$_7$ due to the increase in dimensionality, in agreement with the initial theoretical predictions[16,28]. A prominent derivation from theoretical model should be emphasized that the bandwidth of the band near the Fermi level is observed only about 300 meV from the ARPES experiments[14,15], which is even narrower than that of Sr$_2$IrO$_4$. Reminding us the sharped OCS peaks in the [(SrIrO$_3$)$_m$(SrTiO$_3$)] superlattice with increasing $m$ observed by S.Y. Kim et al. [29]. Both of Nie et al. and S.Y. Kim et al. have been attributed their surprise observations to the band folding due to the additional octahedra rotation about [110] in SrIrO$_3$ besides the rotation about [001] the latter rotation was held in all the RP iridates. Moreover, the Ir-O-Ir bonds along the $c$ direction is essential for the band folding which is therefore lack in the Sr$_2$IrO$_4$ even artificially introducing a [110] rotation in it in calculations [14] . In line to this mechanism, the decease of $m$ reduces the Ir-O-Ir bonds along the $c$ direction in the [(SrIrO$_3$)$_m$(SrTiO$_3$)] superlattices, and would increase the bandwidth. As for the lack of observing sharp $\alpha$ peak in the Moon's experiments of SrIrO$_3$ film grown on MgO substrate[8], S.Y. Kim ascibed that as a result of deffects due to large lattice mismatch[29]. However, a more plausible reason is avialable. The crystal structure of the $SrIrO_3$ film and so its bandwidth is modified seriously by the substrate lattice constraint. According to the crystallographic structures of $SrIrO_3$ films on different substrates[30, 52]. Elongations not only occurs on the lattice constants $a$ and $b$, but also on $c$, at the extension strain side. The lattice angle $\gamma$ is increased as well, introducing a slightly monoclinic distortion in the structures. That is the case for the MgO substrate, enhancement of all the Ir-O-Ir angles therefore would be expected there, leading to increased bandwidth.

At the same time, the varied extent of the orientation dependence suggests potential anisotropic properties in perovskite $SrIrO_3$, which may be investigated by transport measurements with and without magnetic field under different orientation configurations of electrical current and magnetic field. This point claims more investigations on the topic by using small lattice strain film or a single-crystal sample since a recent investigation reported obvious transport anisotropy in $SrIrO_3$ single-layer films (here single-layer means that the film contains only one $SrIrO_3$ layer, not heterostructure with multilayers. It will be abbreviated as SLFs), and suggested that as a result of the epitaxial strain-induced structural distortion[53]. The strong strain dependence of electron transport behavior in $SrIrO_3$ SLFs will be discussed more detail in the section 5.3 of the present review. Another point revealed in the ARPES results is that both the hole pockets and the electron pockets are small, that is, the Fermi level slightly crosses the band edge. The mobility edge of $SrIrO_3$ is therefore quite narrow in accordance with the Mott theory of weak localization, which would cause localization easily [54], in coincident with the experimental observations[12, 45, 55].

## 5. Transport properties of film

This section discusses the transport properties of $SrIrO_3$ SLFs. It is an essential topic for potential spintronics applications, and a strong indicator for determining the electronic band structure as well as its evolution. In line with the general paradigm for transport property experimental investigations, the temperature and the magnetic-field dependence of resistivity have been extensively explored for perovskite $SrIrO_3$ films. The resistivity of the film samples is compatible with that measured in bulk samples in magnitude, and the metal-insulator transition (MIT) is also retained[12, 17, 18, 45, 55] in most cases. At the same time, a large number of behaviors were observed only in the films for example the transport mechanism evolution dependent on film thickness and substrate strain.

**5.1 Basic behaviors**

Representatively, the resistivity of SrIrO$_3$ SLFs, similar to the bulk samples, lightly varies in the range from room temperature to the zero temperature limit [17, 18, 45, 55] (for example see Fig.9a and Fig.12a), even though variations of absolute resistivity exist due to varied microstructure states of the SLFs synthesized by different groups. This behavior is also valid for the SLFs grown on other commonly available substrates such as LSAT, GSO, and NSO [13, 18, 45]. From room temperature to near the MIT point, the SrIrO$_3$ SLFs resistivity was found to be generally proportional to the traditional power law of temperature, that is, $\rho \propto T^p$, with power coefficient $p$ often lower than 2 (see Fig.8a–8c) [13, 17, 45], indicating the non-Fermi liquid behavior[45, 56]. Since $p=3/2$ denotes the inelastic electron–phonon scattering, 1 and 2 denote the inelastic electron–electron scattering, inelastic electron–electron scattering plays a major role in the thin SrIrO$_3$ SLFs, even at high temperature. This is different from bulk perovskite SrIrO$_3$ which demonstrates inelastic electron–phonon scattering associated with $p=3/2$, as shown in Fig.8g [26].

Below the MIT point, the conductance $G$ was proportional to the logarithmic relation of temperature, as seen in Fig.8d–8f [13, 17, 45]. The two-dimensional weak localization is held, so the conductance could be explicitly expressed as [57]

$$G_{2D} = G_0 + q\frac{e^2}{\pi h}\ln\left(\frac{T}{T_0}\right) \quad (1)$$

where $G_0$ is the Drude conductance, $q$ and $T_0$ are constants, $e$ is the electron charge, and $h$ is the Planck constant; $q=3$ for electron–phonon scattering and $q=1$ for electron–electron scattering. Here, the $q$ is varied in a range from about 0.5 to 2.1, implying the carrier is scattered in a mixed mechanism with both the electron–electron scattering and the electron–phonon scattering, but it is likely that the front plays a major role. For bulk perovskite SrIrO$_3$, the low-temperature conductivity follows a linear temperature dependence (Fig.8h), consistent with the three-dimensional weak localization model that is [57]

$$\sigma_{3D} = \sigma_0 + \frac{e^2}{\pi^2 h}\frac{2}{t}T^{\frac{i}{2}} \quad (2)$$

where $t$ is a constant. That $i=2$ hints a mixed carrier scattering, in agreement with that in the SLFs.

We have seen, in perovskite SrIrO$_3$, that the weak localization varies from three-dimensional type in bulk to two-dimensional type in films. It implies the carrier mean free path at low temperature in the SLF is long enough so that the two-dimensional weak localization is still effective in the 35-nm thick film reported by Biswas et al. (except for the film grown on the GdScO$_3$ substrate) [45]. On the other hand, the carrier scattering mechanism is different in bulk and in films, reminding us of the size effect. Biswas et al. investigated the resistance of SrIrO$_3$ SLFs grown on (110)-GSO substrates with diverse thickness (Fig.9a)[45]. Zhang et al. also reported on the transport characteristics of SrIrO$_3$ SLFs on (001)-STO substrates, respectively, with 4-nm and 7-nm thickness[55]. Two most recent preprints studied the density of states evolution around the Fermi level for the SrIrO$_3$ films with precisely unit cell scale thickness by scanning tunneling spectroscopy[58] and ultraviolet photoelectron spectroscopy[59]. The universal trend is that the SrIrO$_3$ films would gradually evolve from metallic to semiconductor/insulator with a reduction in thickness. However, it is not clear the transition is intrinsic due to band structure modification or a result of the enhanced carrier localization at the two dimensional limit.

The weak localization strength is enhanced, as the thickness decreases. For thick films with high quality, the MIT even vanishes (for example the 35nm thick film in Fig.9a). Typically, the insulative regime resistance obeys the simple two-dimensional weak localization transport described by Eq.1 in the films with a relatively larger thickness (Fig.8, thickness ⩾7 nm). For the films with a thickness approaching several nanometers (according to the experiments to date, Fig.9b, the critical thickness could be roughly supposed to be 5 nm), the variable range hopping (VRH) has been the dominant carrier transport mechanism. This mechanism could be formulated as [54]

$$\rho=\rho_0 \exp\left(\frac{T_0}{T}\right)^{\frac{1}{D+1}} \quad (3)$$

where $\rho_0$, $R_0$, $T_0$, and $D$ are constants. The $T_0$ is a function of the effective state density $N$ and the localization length $l$. $D$ (=3, 2) represents the effective dimensionality of the carrier hopping. The hopping mechanism is a signature showing the relatively strong carrier localization, and it could be deemed as the transient regime from the weak localization due to a weak disorder potential to the Anderson localization associated with a full localized carrier. Moreover, the so-called Coulomb gap-induced Efros–Shklovskii VRH (with $D$=1) was also revealed in the ultrathin films at low temperatures[45, 55] (Fig.9b), suggesting a strong electron correlation effect there [55].

So far there are a few reports of the Hall effect of SrIrO$_3$ SLFs. Zhang et al. observed nonlinear Hall effect in the ~7-nm thick films grown on (001)-STO substrates, which provides a transport evidence for the semimetal background state of perovskite SrIrO$_3$ [13] (see Fig.10a). In contrast, the Hall effect for the same thick films grown on (001)-LSAT substrates are nearly linear (Fig.10b). This Hall effect evolution under small in-plane compressive strain suggests that the band structure of SrIrO$_3$ is easily tunable by lattice strain. The tight binding model[16] calculation demonstrated that the suppression of the in-plane hopping of the next-nearest-neighbor $d_{xy}$ orbits, corresponding to the in-plane compressive strain, would distinctly downshift the hole band from the Fermi level[13]. For thick SIO SLFs, the nonlinear feature of Hall traces was usually indistinct [18, 53], perhaps because the perovskite structure is partially relaxed to the monoclinic structure in thick SIO films. L. Fruchter recently noticed in thin SIO SLF that a nearly linear Hall effect could be observed at a temperature above the metal-insulator transition point even through the nonlinear Hall effect was also demonstrated at low temperature. A nominal Hall coefficient for the nearly linear Hall traces could also be extracted and was shown to be not strongly temperature dependent above the MIT temperature but slightly increased as temperature decrease in the weak localization regime, indicating the reduction of carrier density due to localization[53]. The Hall traces provided the carrier density at a magnitude order of $10^{19}$ cm$^{-3}$, and the carrier mobility at a magnitude order of $10^2$ cm$^2$ V$^{-1}$ s$^{-1}$, according to the two-carrier transport model for isotropic materials[13].

**5.2 Magnetotransport**

Typically, SrIrO$_3$ SLFs exhibit a "W" type magnetoconductance trace (MC), that is, a negative MC at low field region and positive MC at proper high field region (see Fig.10c and 10d). Generally, the high field positive MC, corresponding to a negative magnetoresistance (MR), is a transport signature of broken weak localization and is taken as the MR background. The low field negative MC is a result of the strong SOC in iridates and is regarded as a quantum correction to the positive MC background. From the viewpoint of spin splitting, SOC is equivalent to a Zeeman field, so it can break the time reversal symmetry of the two paths with an opposite electron moving direction, leading to the crash of the electron self-interference premise for weak localization. This effect is called the *weak anti-localization* [60]. Besides the familiar negative MR background, pure positive MR was also occasionally observed in the SrIrO$_3$ SLFs, which is due to the absence of MIT[45], and further proving the disorder-associated weak-localization mechanism of the MIT and the negative MR.

The MC curve of the SrIrO$_3$ SLFs can be frequently analyzed using the Hikami–

Larkin–Nagaoka equation (HLN model) [55, 61] for the cases with weak anti-localization superposed on the weak localization effect suppressed by magntic field, that is

$$\frac{\Delta G_{SOC+WL}}{G_u} \propto \psi\left(\frac{1}{2}+\frac{B_e}{B}\right)-\psi\left(\frac{1}{2}+\frac{B_i+B_{soc}}{B}\right)+\frac{1}{2}\left[\psi\left(\frac{1}{2}+\frac{B_i}{B}\right)-\psi\left(\frac{1}{2}+\frac{B_i+2B_{soc}}{B}\right)\right] \quad (4)$$

with $\psi(x)$ as the digamma function, $\Delta G = G(B)-G(B=0)$ and $G_u=e^2/(2\pi^2\hbar)\approx 1.2\times10^{-5}$ S. The $B_e$, $B_i$, and $B_{soc}$ are the equivalent fields for elastic scattering, inelastic scattering, and the scattering induced by SOC, respectively. All of them can be formulated by $B_o=\hbar/4e\,l_o^2$ with their scattering lengths $l_o$ (o=e, i and soc). In the ultrathin SrIrO$_3$ SLF, the electron–electron interaction (EEI) is non-negligible, and a parabolic field-dependent MC modification should be further considered [55], i.e.,

$$\frac{\Delta G_{SOC+WL+EEI}}{G_u} \propto \psi\left(\frac{1}{2}+\frac{B_e}{B}\right)-\psi\left(\frac{1}{2}+\frac{B_i+B_{soc}}{B}\right)+\frac{1}{2}\left[\psi\left(\frac{1}{2}+\frac{B_i}{B}\right)-\psi\left(\frac{1}{2}+\frac{B_i+2B_{soc}}{B}\right)\right]+\gamma B^2$$

(5)

Based on the MC curve fit using the HLN model, Zhang et al. discovered that the SOC (indicated by $B_{soc}$ and the corresponding Rashba coefficient $\alpha$ in the SrIrO$_3$ SLFs) is enhanced with temperature (see Fig.11)[55], deviating from the traditional prediction of the k.p theory where SOC is independent of temperature. The derived Rashba SOC coefficient $\alpha$ increases about 30–45% per Kelvin, about 100 times over the rate of typical semiconductor quantum wells [62]. This temperature-sensitive behavior suggests a novel feature of strong SOC, and it can be thoroughly explained by a relation with linearly temperature-dependent Lander g factor, $g=g_0+\lambda T$, [55]

$$\alpha = \left[g_0(1-g_0)+(1-2g_0)\lambda T-\lambda^2 T^2\right]\frac{\pi e\hbar^2 \varepsilon}{4m_*^2 c^2} \quad (6)$$

where $\varepsilon$ is the asymmetric structure-induced electric field, $c$ is the light speed, $g_0$ is the zero temperature Lander g factor, $\lambda$ is the linear temperature coefficient of the Lander g factor, $\hbar$ is the reduced Plank constant, and $m_*$ is the effective electron mass.

Recently, L. Fruchter et al. proposed an alternative understanding of the MC data under different temperatures through a fitting model adding a temperature-dependent magnetic scattering term , so the $B_{soc}$ could be set as temperature independent in the fitting of the MC. The biggest difficulty of the interpretation is what could be the magnetic scattering source since SIO films with thickness above ~4 unit cells are paramagnetic [19]. L. Fruchter et al. supposed the source as paramagnetic scattering centers with a Curie-like effective moment[53]. In addition, the fitted magnetic scattering strength is unreasonable as a divergence as $T\to 0$, as pointed out by the authors. The fitted magnetic scattering strength demonstrated a minimum at a definite temperature, which is also unexpected.

Now we would like to outline the basic transport features of $SrIrO_3$ SLFs, 1) non-Fermi liquid behavior was often observed; 2) weak localization associated metal insulator transition; 3) weak antilocalization effect induced by strong SOC, which causes the positive magnetoresistance at low field on a negative MR background corresponding to the broken weak localization; 4) weak nonlinear Hall effect generated by the electron carrier dominated semimetallic state; 5) prominent size effect caused metal-insulator transition with thickness reduction.

### 5.3 Strain effect

In Section 5.1, it was shown that in-plane compressive lattice strain could suppress the nonlinearity of the Hall traces of $SrIrO_3$ films. Another interesting evidence of the sensitive strain effect is the recently noted RT slope jumping at the STO substrate phase transition point from cubic to tetragonal[40, 53]. For film systems, a substrate lattice mismatch strain puts two essential effects on film, changing the crystal structure (lattice constant with $IrO_6$ octahedra configuration) and introducing defects. The first one could intrinsically modify the band structure and the latter will affect the carrier scattering.

Carefully experimental determination of the crystallographic structures of $SrIrO_3$ SLFs on varied substrates has been reported most recently[52]. It made quite surprised results summarized in Fig.12. The lattice constant $a$ and $b$ are increased in both of the extension side and the compressive side, and $c$ is elongated by extension but shortened by compressive strain (Fig.12b). Accompany, the $\gamma$ angle is enlarged by extension and reduced at the compressive side (Fig.12a), with the other two unit cell angles kept at 90°, transforming the space group to $P112_1/m$[30]. This lowering of symmetry was proved to lift the Dirac node of $SrIrO_3$ in a calculation stated by Liu et al.[30].

In $SrIrO_3$ SLFs, from the viewpoint of conduction (Figs. 13a-13c), the resistivity of $SrIrO_3$ SLFs on (001)-STO substrates are commonly lower than those of SLFs grown on other frequently adopted substrates, even the in-plane lattice mismatch strain of (001)-STO substrate is larger than that of the (110)-DSO and the (110)-GSO substrates. When taking the resistivity of the films on the (001)-STO substrates and the in-plane lattice constant of the STO substrate as references, the resistivity of the films under different substrate strains ranging from compressive to tensile form a "V" shaped tendency around the zero point of lattice mismatch defined by STO substrate lattice constant as the reference (see Fig.13c). Moreover, the resistivity is linearly dependent on the lattice mismatch on both the negative and the positive side, and the slopes of the two sides are comparable (Here it is unusual that there is a downward tendency of resistivity for the films on DSO, GSO and NSO substrates at low temperature. The slopes at 6K would be closer if using the resistivity data naturally obtained by extrapolating the curves at the temperature regime above the downward tendency to 6K, according to the resistivity curve tendency like that in the films on STO substrates). The carrier density similarly obeys an inverse V-shaped trend around the STO substrate (see Fig.13e). At the extension strain side, we have seen all the lattice constants $a$, $b$ and $c$ were elongated in SrIrO$_3$ films [52], which would expand all the Ir-O-Ir angles and so the bandwidth. The decrease of carrier density is therefore understandable. At the compressive side, the shrinkage of $c$ in contrast to the expanding of $a$ and $b$ [52] makes it complex. In the starting model of the $J_{eff}$=1/2 state in iridates with an imagined cubic environment, the $d_{xy}$, $d_{xz}$ and $d_{zy}$ orbitals are mixed equally[7]. However, the $J_{eff}$=1/2 state is fragile[28]. The complex IrO$_6$ octahedral rotation will wipe the equality. Compressive strain elevates the $d_{xy}$ orbital level higher in energy than the $d_{xz}$ and the $d_{zy}$ levels[63], inducing more occupation in the latter two orbits in SrIrO$_3$. Consequently, the shrinkage of $c$ would decrease the hopping along the two orbitals. In contrast, the carrier mobility bears a monotonously decreasing tendency with the strain evolving from compressive to tensile (Fig. 13d).

Generally, the structure and transport behavior evolution due to strain effect have demonstrated that the band structure and associated physical properties of SrIrO$_3$ film are highly tunable. However up to now, we still know less about the details of these tunable features in experiments, much understanding are theoretical or just in experience. The existing experiments are inadequate on both of the used substrates to supply varied strains and the probed physics effects, which call more investigations on the varied physics with strain and/or high pressure stimulus.

## 6. Magnetic properties

Unlike most iridates, bulk perovskite SrIrO$_3$ is paramagnetic. This state was retained in thick films. For ultrathin SLFs, the result is not clear. In principle, we can roughly predict that the band structure of SrIrO$_3$ film would approach the band structure of Sr$_2$IrO$_4$ as the thickness decreases, so a similar magnetic state is perhaps generated. However, the magnetic state features in ultrathin SrIrO$_3$ SLFs are too weak to be detected definitively, just as was observed by the authors of this review.

Matsuno *et al.* adopted the [(SrIrO$_3$)$_m$/SrTiO$_3$]$_k$ ($m$ = 1, 2, 3) superlattice structures like shown in Fig.14a to increase the magnetic signals, and detected obvious magnetic ordering transition signatures in RT, MT, and MH (see Fig.14b-14e), where $m$ indexes the unit cells of every SrIrO$_3$ layer and $k$ is the periodic number [19]. An abrupt resistance enhancement induced by magnetic transition, similar to what we have seen in the ultrathin SrIrO$_3$ SLFs, was demonstrated in the RT curves of the $m$=1 and 2 superlattices. The transition temperature was decreased with an increase in $m$, see Fig.14b. Corresponding transition points were displayed in the MT traces (Fig.14d). Below the transition points, abruptly increased Hall coefficients were also observed.

The MH loop of the $m$=1 superlattice shows a coercive field about 0.02T, and a ferromagnetic ground state with a saturated moment about 0.02$\mu_B$/Ir. Further MT with the loaded magnetic field parallel or perpendicular to the IrO$_2$ plane revealed that the weak ferromagnetic-type magnetic ordering is only constructed in the parallel case (Fig.14f). The resonant magnetic X-ray diffraction at the Ir $L_3$ edge of the m=1 sample where only the in-plane antiferromagnetic ordering corresponding to the σ-π' polarization peak at (0.5, 0.5, 5) was manifested (Fig.14g) [19]. Therefore, it is believed that the weak moments are a result of the Dzyaloshinskii–Moriya (DM) interaction associated with the rotation of the IrO$_6$ octahedra, similar to the Sr$_2$IrO$_4$. The IrO$_2$ interlayer coupling is a ferromagnetic type, and an antiferromagnetic (AFM) structure is formed in the IrO$_2$ plane, schematically depicted in Fig.14a.

SrIrO$_3$ has a larger lattice constant compared to that of Sr$_2$IrO$_4$, corresponding to a larger Ir-O-Ir angle, therefore there is a smaller IrO$_6$ rotation, which results in a weaker DM interaction. At the ultrathin limit of the [(SrIrO$_3$)$_m$/SrTiO$_3$]$_k$ superlattice, due to the in-plane compressive strain constraint supplied by the STO layer, the IrO$_6$ octahedra are rotated more and more as decreasing the stacking IrO$_2$ planes $m$, resulting in stronger magnetic coupling and electron correlation [19]. Reducing $m$ simultaneously depresses the interlayer coupling along *c*-axis. From this point of view, $m$ is smaller and the superlattice is more similar to Sr$_2$IrO$_4$. This was also proven by the band structure simulation results of the $m$=1 and 2 superlattices where a J$_{eff}$=1/2 band gap was gradually opened with a decrease in $m$ when the electron correlation energy $U$ is set at high value, as 3eV [19]. More calculations with high $U$ reveled that canted AFM order is founded with compressive strain just as experiment demonstrated while tensile strain favors the collinear AFM state[63].

The discussions thus far regarding magnetic ordering induced in ultrathin SrIrO$_3$ systems [19] have been fairly phenomenological. A clear microscopic model and experimental results to depict the generation and, in particular, the evolution of a detailed spin configuration with the stacking number of $IrO_6$ octahedral layers are still lacking. Besides, we could note that a high $U$ is requested for achieving the AFM state in $SrIrO_3$ in all the calculations to date[16, 19, 51, 63]. However, it is still unclear how the high $U$ requested is founded in the few unit cell thick $SrIrO_3$ layers? Experimentally, the OCS results on same $[(SrIrO_3)_m/SrTiO_3]$ superlattice samples[29] discussed in section 4 showed much wider excitation peaks in the $m$=1 and 2 samples, which in principle implies wider bandwidth and so weaker effective $U$. $U$ is not necessary for depicting the band structure of $SrIrO_3$ observed by ARPES[14, 15]. The large $U$ cannot be originated from the intrinsic electron correlation in $SrIrO_3$, because the $U$ is only about 0.5eV in iridates in the present prevail understanding. Bacially, high $U$ is for openning the gap between LHB and UHB, so realizing a band structure same to that of $Sr_2IrO_4$. B.J. Kim's pointed out that the magnetic ordering would be attributed to a reduction of the intralayer exchange interaction with decreasing the interlayer distance by strain or by reducing the $IrO_2$ layers ($m$ in the $[(SrIrO_3)_m/SrTiO_3]$ superlattice)[51]. Therefore, not only the Ir-O-Ir angles but also the Ir-O bonds length was taking into account in their fully *ab initio* calculations with combining DFT and the constraint random phase approximation (cRPA) method. It was predicted that the $U$ is increased with m decrease, approaching 1.59eV at $m$=1. We have seen that the substrate stain induced symmetry lowering from the *Pbnm* to the $P112_1/m$ [30, 52]. That supplies a possible to achieve the request high $U$ or other mechanism for magnetic ordering. This picture reminds us that it is possible to detect incontrovertible evidence of magnetic ordering behavior in $SrIrO_3$ SLFs grown on a substrate with a greater compressive strain than the $SrTiO_3$ substrate. More studies on the structure and magnetic properties of the $SrIrO_3$ superlattices and more calculations on the magnetic properties of $SrIrO_3$ with the $P112_1/m$ structure distortion are in demanding in future. Additionally, whether possible to see a $SrIrO_3$-layer exchange coupling tuned by the thickness of the interlayer is also an interesting topic.

## 7. Topological phase predictions based on the $SrIrO_3$ film system

The most attractive feature of perovskite $SrIrO_3$ is the Dirac node crossed by the Fermi level (see Fig.1b and 1c), first proposed under the frame of a tight-binding Hamiltonian constructed by Carter et al. based on the $J_{eff}$ =1/2 single band model [16]. This Dirac node hosts the electron carriers for the semimetal background state of SIO and can evolve into different kinds of topological states depending on the specific symmetry breaking[16, 20-22], mainly proposed by the work produced by Professor Kee's group. This has the potential for opening up a new field of topological electron structure materials, that is, perovskite oxide topological matter.

The unit cell for perovskite SIO shown in Fig.1a can be considered to comprise two same parts with same rotation $\pi$ with respect to each other, where the two parts share the $Ir^{4+}$ plane at $Z=c/2$ [20]. The *Pbnm* space group ($\hat{\Pi}_m$) determines that the tight-binding Hamiltonian[16] defined on it and the $J_{eff}=1/2$ spin space has a mirror symmetry and a chiral symmetry [20]. The Dirac node is generated from the mirror symmetry, and robust under a moderate Hubbard interaction [16]. It is expanded into a nodal ring as a result of the Fermi level shifting away from the Dirac point, due to the attached orbit hopping coming from the $IrO_6$ octahedra rotation and tilting [16, 20]. Consequently, the size of the nodal ring could be tuned by the octahedral status [20]. The nodal ring is manifested as a non-trivial localized and dispersionless zero-energy state which is characterized by a pair of integers with topological one-dimensional winding number along the $k_c=\pi/c$ line on the (110) and the ($1\bar{1}0$) surface. This is protected by both the mirror and chiral symmetries determined by the crystal structure, therefore providing the so-called topological crystalline metal [20].

The nodal ring can evolve into various topological surface states by breaking the mirror symmetry while keeping the chiral symmetry. This was proposed as three kinds of cases. **1)** A [110]-direction magnetic field would lift the nodal ring as a pair of three-dimensional nodal points which would be further split into four Weyl nodes due to the sublattice interactions [20]. **2)** A [$1\bar{1}0$]-direction magnetic field would shift the zero-energy state from the $k_c=\pi/c$ line to the $k_c=\pi/c \pm \delta$ ($|\delta| \ll 1$) line on the ($1\bar{1}0$) surface [20]. **3)** If introducing a staggered sublayer interaction, which could be realized through the superlattice along the c axis, the nodal ring would shrink into a pair of three-dimensional nodal points under a weak mirror-symmetry breaking because the degeneracy of the nodal ring was lifted except at the two nodal points. However, a trivial insulator state would come forth when the mirror symmetry was strongly broken. Between the two states, that is, where the mirror symmetry was moderately broken, there exists a strong topological phase [16]. This method suggests superlattice candidates constructed by $SrIrO_3$ and other proper orthorhombic oxides such as $SrRhO_3$ [16], $SrCoO_3$ [16], $SrZrO_3$ [22], $SrHfO_3$ [22], and $CaTiO_3$ [22]. On the same basis, the superlattice heterostructure $(SrIrO_3)_{2m}/(CaIrO_3)_{2n}$ was proposed as a non-symmorphic

Dirac semimetal with double Fermi arcs on the (001) plane, and is thus the first candidate to demonstrate helicoid surface states [64]. The particular requirements for these superlattices rest on the SrIrO$_3$ layer being at an ultrathin limit (single UC are necessary for most of the cases) [16, 22] and perfect interfaces are desired, presenting a significant challenge for experimentally verifying these supposed topological states.

Experimentally, ARPES is apparently the most effective method to verify these particular topological surface states. However, unfortunately, so far no relevant experiment has been reported. This probably stems from the challenge of growing atomic scale, flat and smooth, single IrO$_6$ octahedral layers, which absolutely require atomic scale, flat and smooth substrates and a highly delicate tuning of the growth conditions. On the other hand, the predictions for the topological states are highly dependent on the original *Pbnm* structure symmetry and the validity of the tight-binding model constructed on the basis of the $J_{eff}=1/2$ single-band model as pointed out by Ref.[21]. It has been confirmed from both theoretical calculations [21, 28] and experiments [14, 15] that the $J_{eff}=1/2$ band is partially mixed with the $J_{eff}=3/2$ band in SrIrO$_3$. In addition, the breaking of *Pbnm* symmetry due to substrate strain[30] also needs to be considered.

In addition, an anomalous quantum Hall effect and fractional quantum Hall effect were also theoretically proposed in the superlattice system stacked by perovskite SrIrO$_3$ and SrTiO$_3$ grown along the [111] direction, based on the half-filled and nearly flat band generated by the Haldane-type honeycomb lattice on the interface [5, 6, 65]. Lado et al. suggested the realization of topological semimetal in a quite particular multilayers (SrTiO$_3$)$_7$/(SrIrO$_3$)$_2$ stacking along [111][66]. A recent work has reported that it is possible to grow high-quality metastable ultrathin perovskite SrIrO$_3$ film on (001)-SrTiO$_3$ substrate[67], which constitutes the first step.

## 8. Conclusions

As a particular metallic member in the strong spin–orbit coupling 5d iridates, orthorhombic SrIrO$_3$ has attracted considerable interest recently. The experimental research to date has revealed its basic physical properties as a correlated semimetal with strong spin–orbit coupling, holding great promise for achieving a novel oxide topological matter. Orthorhombic SrIrO$_3$ is located at the boundaries of versatile ground states. The configuration of the IrO$_6$ octahedra, for example, the rotation and tilting angles, is essential to determine the ground states of perovskite SrIrO$_3$, while simultaneously SrIrO$_3$ has another monoclinic structure phase. Altogether, this gives us the opportunities and routes to tune $SrIrO_3$ into various states; thus, we can explore the physical properties under this competition and among multiple interactions including spin–orbit coupling, crystal field, and electron correlations, as well as how these interactions determine the electron and magnetic structures.

The band structure of $SrIrO_3$ films under varied strain states and varied dimensionality (number of $IrO_2$ layers) are not clear, and in high request, which is vital for understanding the transport and magnetic evolution with strain and dimensionality. The topologically protected Dirac nodal ring in perovskite $SrIrO_3$ is still an open question. It bears real significance because the nodal ring is not only the starting point of all the supposed topological phases but also a probe into the validity of the so far prevalently accepted Hamiltonian models describing perovskite $SrIrO_3$. Besides the supposed superlattice systems, we suggest the fabrication of $SrIrO_3$ film without strain by design a substrate with buffer layer so with lattice constants precisely matching with those of $SrIrO_3$. We also suggest a gate experiment to tune the Fermi level sweeping the Dirac node, which is possible since the detected carrier density of the SLF is at the gate tunable boundary.

Moreover, the recent research is still mainly focused on the $SrIrO_3$ itself, not extended to its heterostructure/multilayer. The strong spin–orbit coupling in $SrIrO_3$ might affect the physical properties of other perovskite layers. It can be expected that research of superlattice systems composed of the $SrIrO_3$ layer should demonstrate a series of novel physical properties like the magnetic state in $SrIrO_3/SrTiO_3$ superlattices [47]. Several close recently published works have reported on $SrIrO_3$ dimensionality controlling a magnetic easy axis reorientation in $La_{2/3}Sr_{1/3}MnO_3/SrIrO_3$ superlattices[68], spin-glass-like behavior and a topological Hall effect in the $SrRuO_3/SrIrO_3$ superlattices/bilayer [69, 70], as well as emergent ferromagnetism in $SrMnO_3/SrIrO_3$ superlattices [71]. Generally, due to spin–orbit coupling, in principle, tuning the spin state of charges, research into superlattice systems composed of a $SrIrO_3$ layer and an alternating layer such as multiferroics materials, superconductors, and topological phase materials is definitely expected. Additionally, the $SrIrO_3$ film was discovered recently exhibiting excellent electrocatalyst performance in the oxygen evolution reaction[72, 73] due to the iridium ions, which perhaps will excite a research field for iridates in the near future.

**Acknowledgment:** This work was financially supported by the National Natural Science Foundation of China (Grant Nos. 51472112, 51032003, 11374140, 11374149,

10974083, 11004094，11134006, 11474150 and 11174127), and the National Basic Research Development Program of China (973 Program) (Grant Nos. 2015CB921203, 2014CB921103, and 2015CB659400). L.Y Zhang also acknowledges the financial support of the National Natural Science Foundation of China (Grant No. 51402149) and the Fundamental Research Funds for the Central Universities (Grant No. 20620140630).

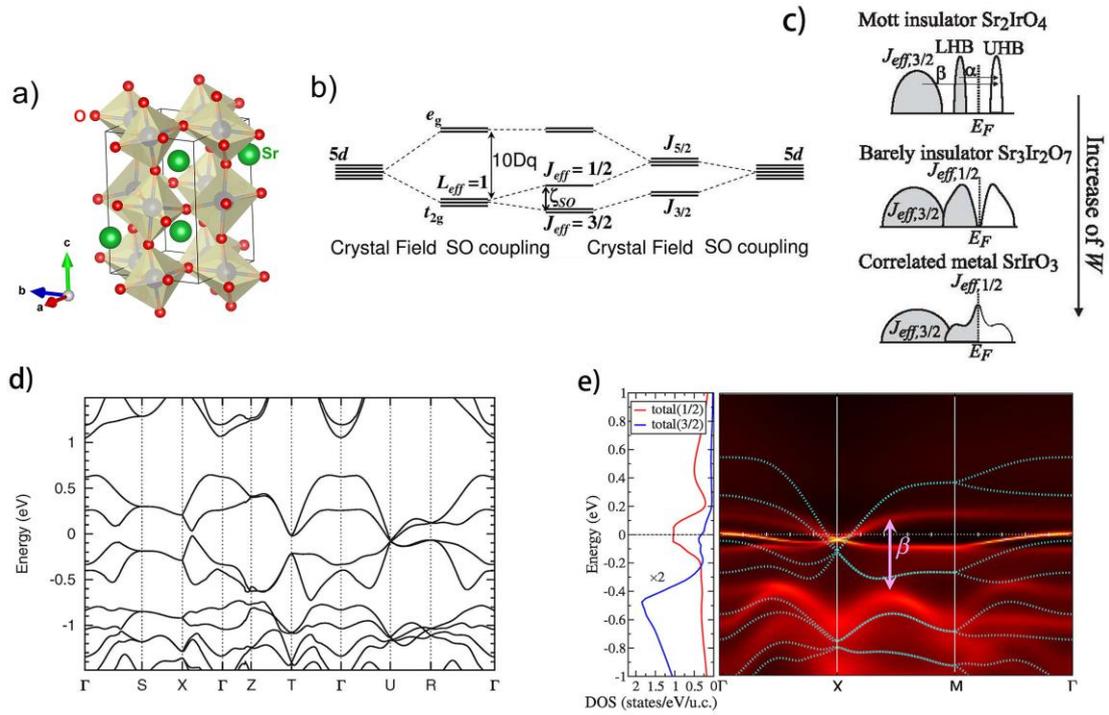

**Fig.1 a)** Crystal structure and **b–e)** electronic structure of orthorhombic $SrIrO_3$. b) Orbit configuration of $5d^5$ with SOC and $U$ [7]; c) schematic picture of the band evolution with dimensionality increase in $Sr_{n+1}Ir_nO_{3n+1}$ iridates. $W$ represents the bandwidth [8]. the α notes the excitation of electrons from LHB to UHB, and the $β$ for the excitation the $J_{eff}$=3/2 band to the UHB seen from the optical conductivity spectra [8, 29]; d) calculated with the LDA+SOC+$U$ method [16] and e) with the LDA+DMFT method [28]. The dotted lines are obtained by the generalized gradient approximation plus $U$ (GGA +$U$) method, and the background bright curves correspond to the calculated results through the LDA+DMFT.

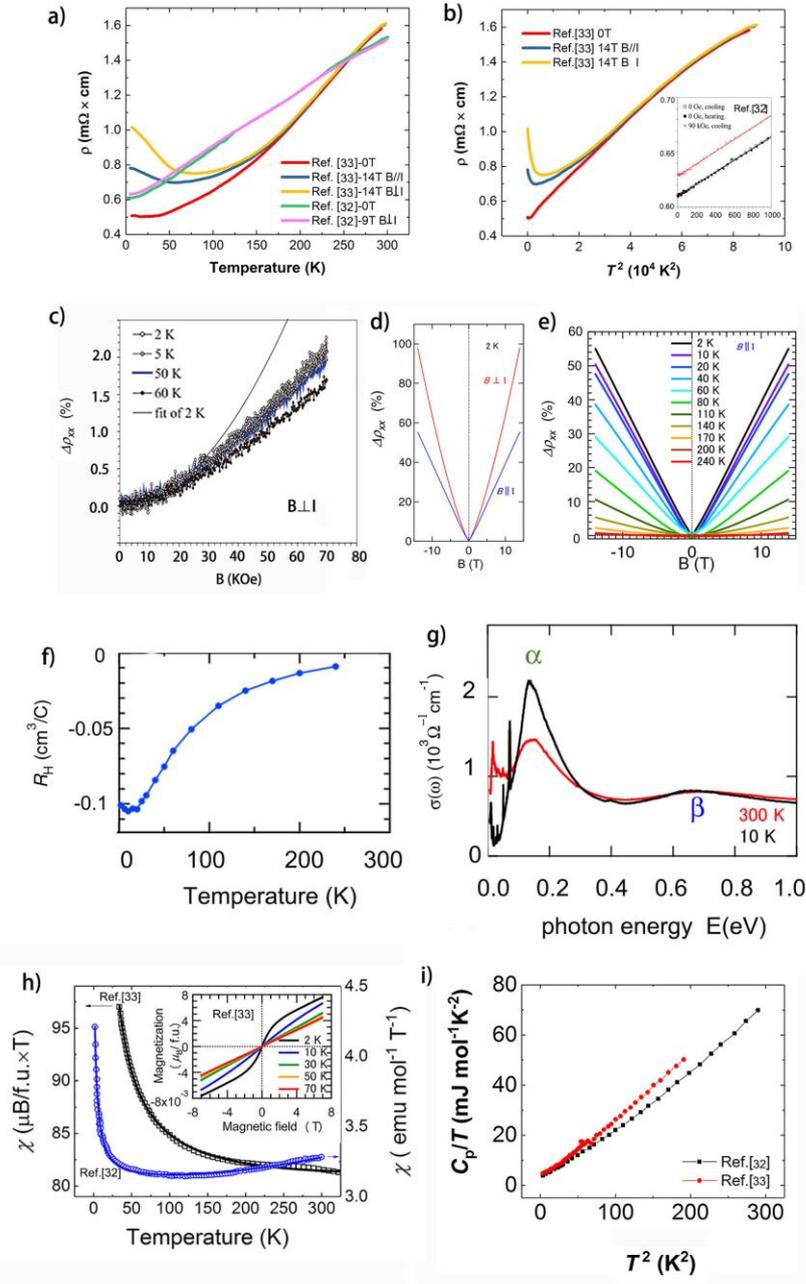

**Fig.2** Electronic transport and magnetic properties of a polycrystalline orthorhombic $SrIrO_3$[32, 33]. **a)** Temperature dependence of resistivity; **b)** shows the Fermi-liquid type $T^2$ dependence of the resistivity; c) Magnetoresistance at varied temperatures of the sample reported by Ref.[32]; **d)** Magnetoresistance at 2K of the sample reported by Ref.[33]; e) Magnetoresistance at varied temperatures of the sample reported by Ref.[33], with magnetic field parallel to current. f) Temperature dependent Hall coefficient of the sample in Ref.[33]; g) optical conductance spectrum of the sample in Ref.[33]; h) Magnetization susceptibilities, inset shows the MH loops at varied temperatures; i) temperature-dependent specific heat

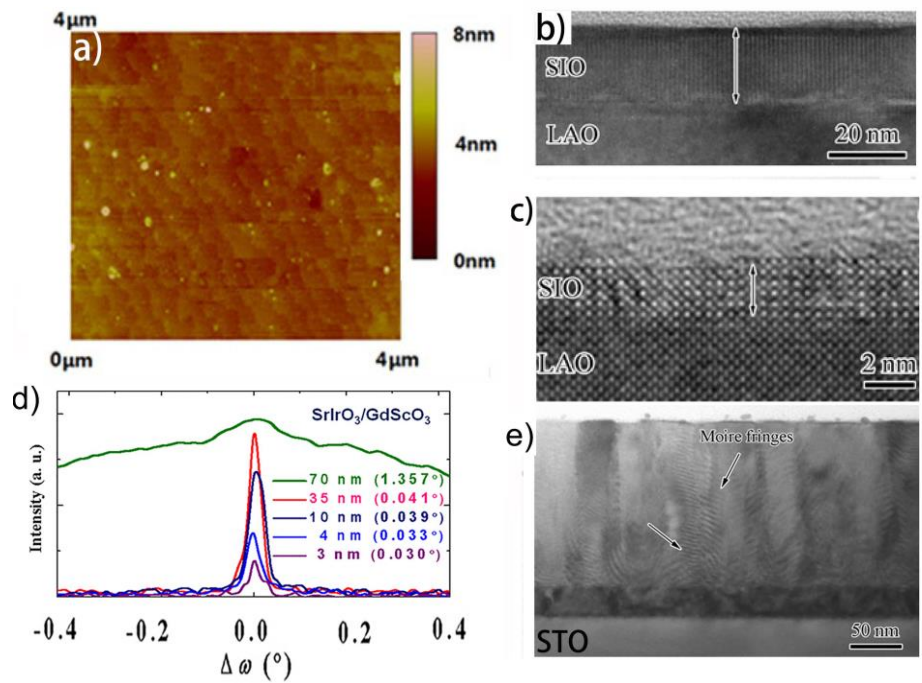

**Fig.3** Microstructures of SrIrO$_3$ films. **a)** AFM-imaged surface steps showing the two-dimensional growth model of thin films[12]; **b) and c)** HRTEM images showing the atomic epitaxial growth of the thin films on (001)-LaAlO$_3$ substrates[12]; d) Rocking curves around the (001)$_C$ peak of SrIrO$_3$ films of various thicknesses on GdScO$_3$ substrates[45]; **e)** TEM contrast profile of a 140-nm thick film grown on the (001)-SrTiO$_3$ substrate. A strain-confined 30-nm thick layer with orthorhombic phase and strain-relaxed layer indicated by the Moiré fringes were distinctly demonstrated [24].

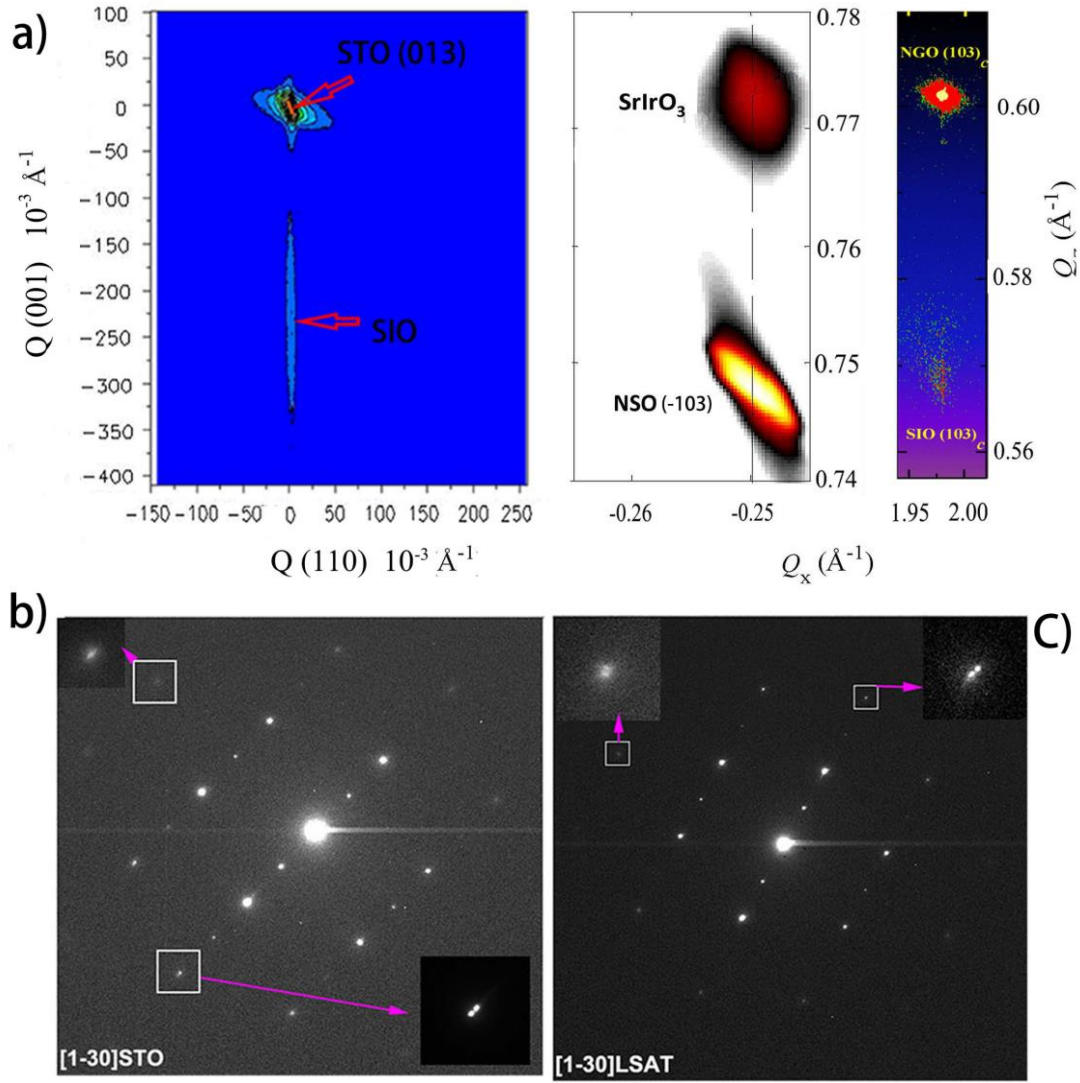

**Fig.4 a)** X-ray reciprocal space mapping (RSM) images of SrIrO$_3$ films grown on (001)-SrTiO$_3$ [13], (001)-NbScO$_3$ [18], and (110)-NdGaO$_3$ [45] substrates. Selected-area electron diffraction (SAED) patterns of the 7nm thick SrIrO$_3$ films on the **b)** (001)-SrTiO$_3$ and **c)** the (001)-LSAT substrates, where the diffraction spots along the out-plane direction are split but not along the in-plane direction, suggesting a fully constrained in-plane lattice [13].

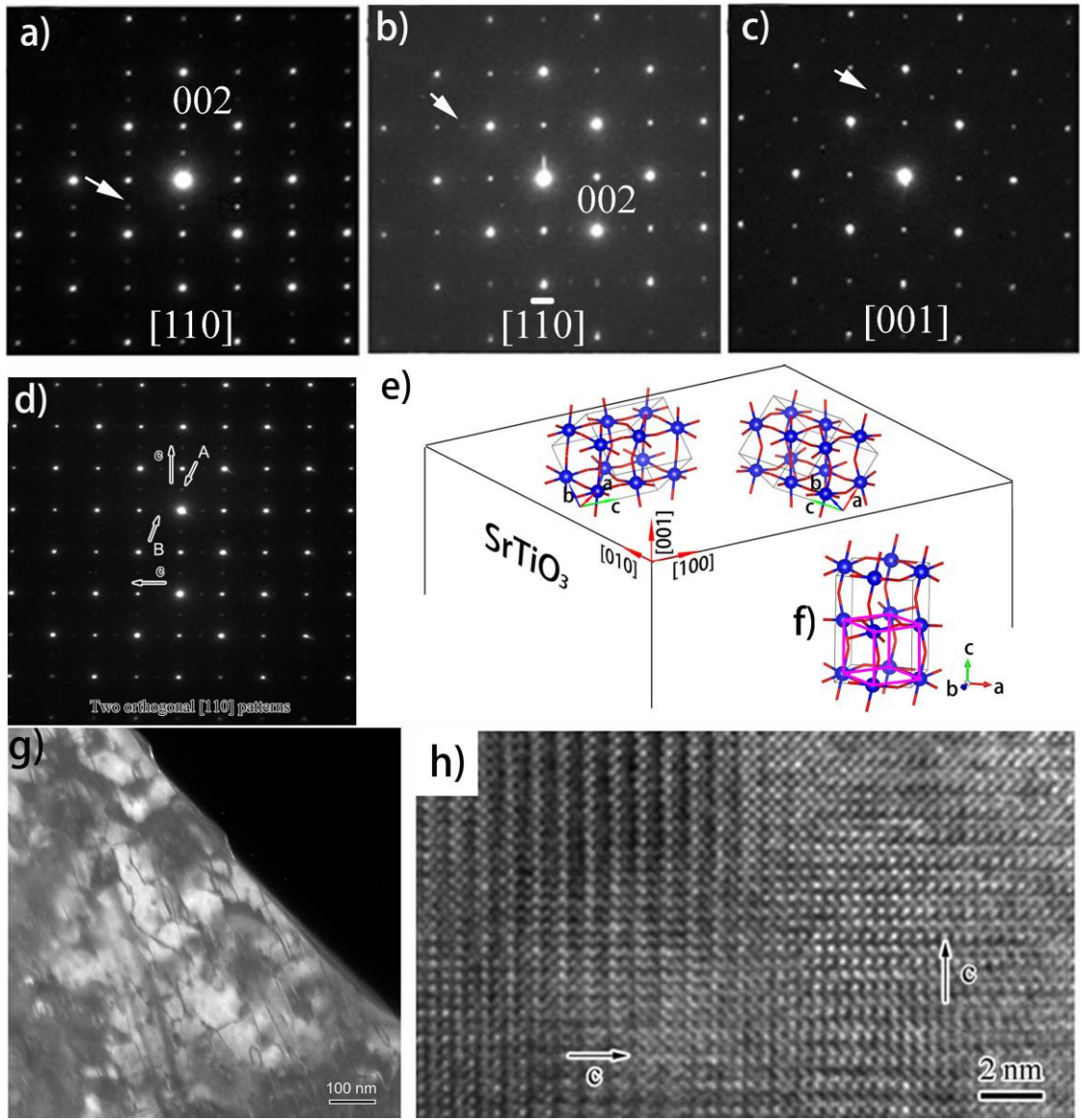

**Fig.5 a)–d)** SAED patterns of a 30-nm thick SrIrO$_3$ film on a (001)-SrTiO$_3$ substrate from three assumed orthogonal axes, indicating the orthorhombic crystal structure of the film. The arrows point to the spots corresponding to two types of domains shown as g) and h). **e)** Schematic depiction of the epitaxial growth orientation matching the substrate of the two types of domains. **f)** Schematic demonstrating the pseudo-cubic structure of the orthorhombic SrIrO$_3$. **g)** Antiphase boundaries of the two types of domains and **h)** the HRTEM showing the two types of domains rotated with respect to each other at 90 °. [24]

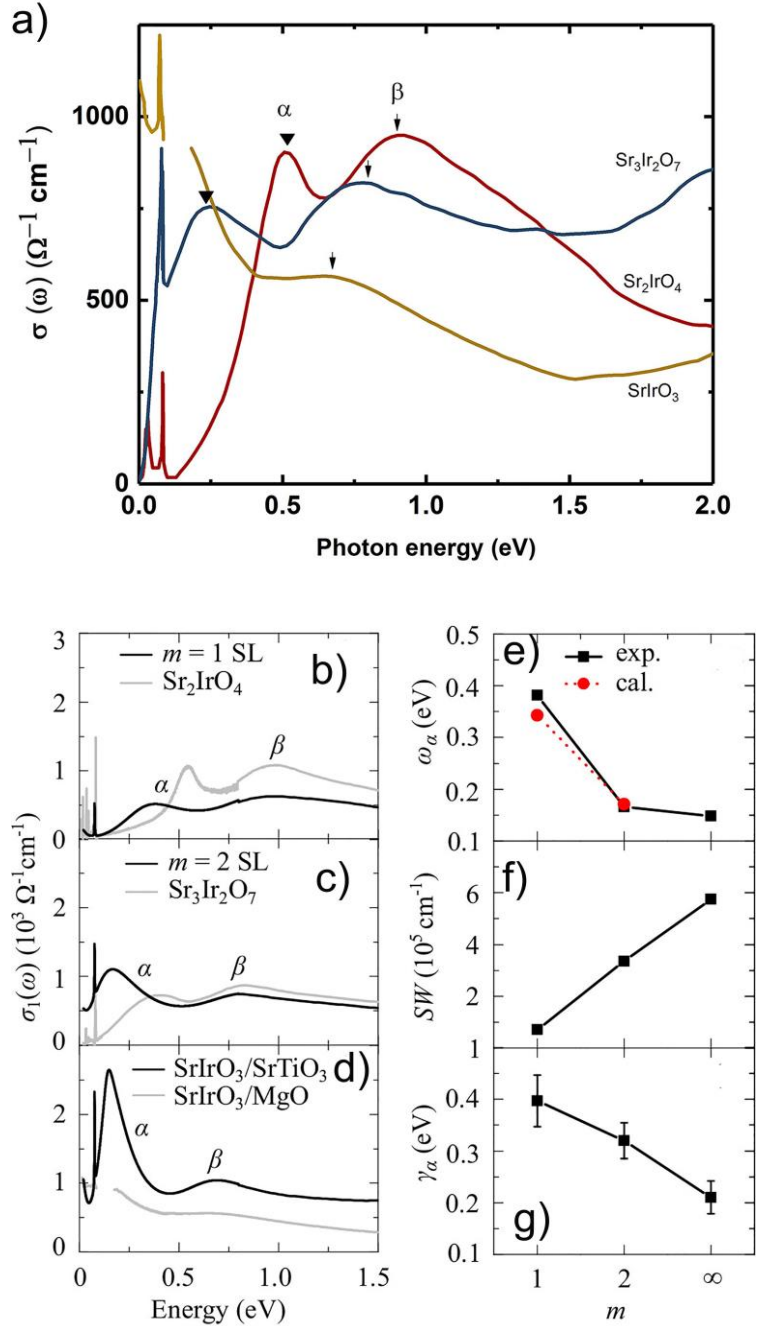

**Fig.6.** Optical conductivity spectra of $SrIrO_3$ films. **a)** the real part $\sigma_1(\omega)$ of optical conductivity at room temperature of a $SrIrO_3$ film grown on MgO substrate, in comparison with $Sr_2IrO_4$ single crystal and $Sr_3Ir_2O_7$ single crystal[8]. **b)-d)** the real part of optical conductivity $\sigma_1(\omega)$ at 20 K of superlattice $[(SrIrO_3)_m(SrTiO_3)]$ with b) $m=1$, c) $m=2$ and d) $m=\infty$ (a $SrIrO_3$ film grown on STO substrate). **e)** The position of the $\alpha$ peaks. Black squares indicate the experimental data, and the DFT+$U$ calculation data are plotted as red circles. **f)** The low energy spectra weight obtained by integrating $\sigma_1(\omega)$ from 20–37 meV. **g)** The widths of the $\alpha$ peaks of the superlattice samples in b)-d). Figure b)-g) were adapted from Ref [29]

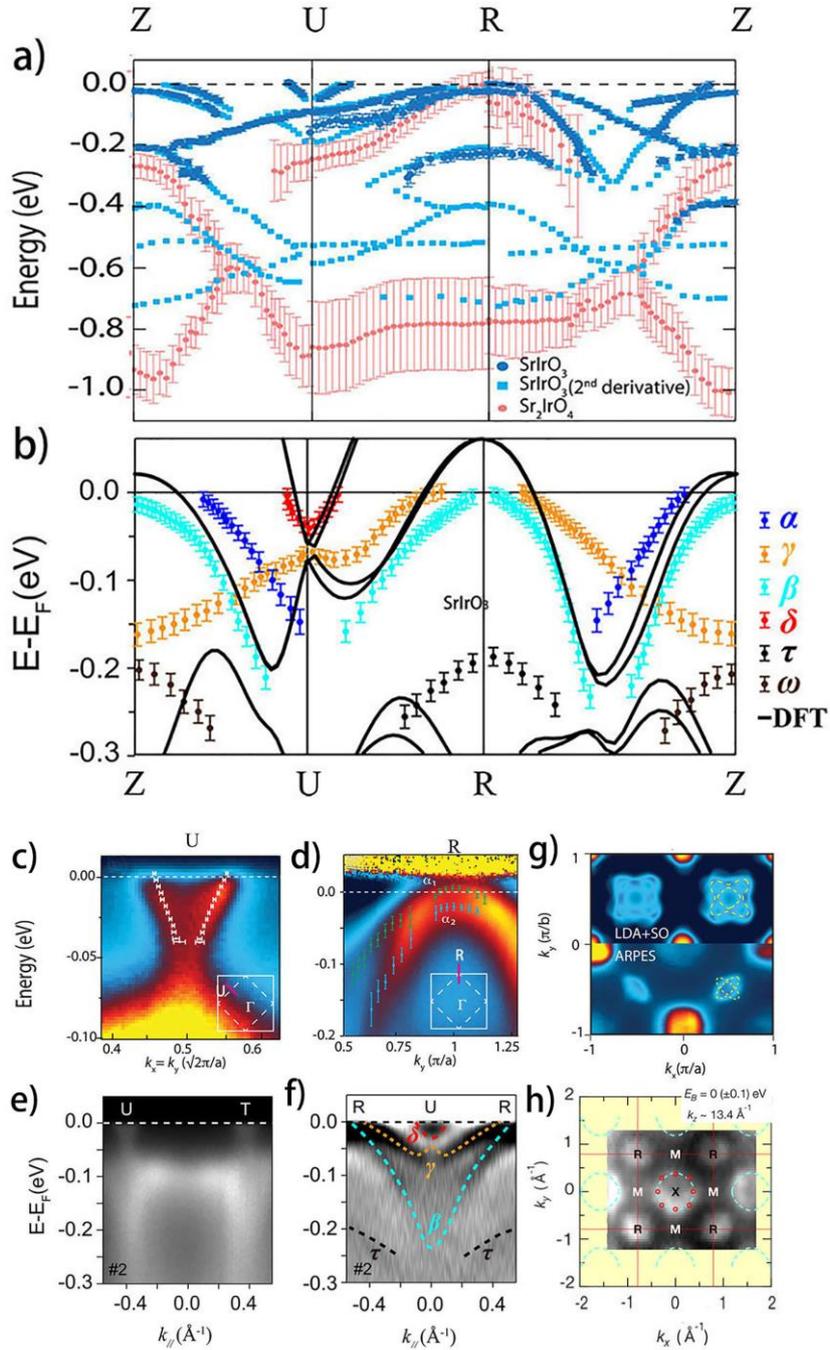

**Fig.7** ARPES-imaged band structure of $SrIrO_3$ films. a) band structure of a film grown on a (001)-LSAT substrate[14], with c) and d) as the corresponding band at specific points U and R, respectively. They show the electron pocket and hole pocket, respectively. g) shows the corresponding measured and calculated Fermi-surface. b) band structure of a film grown on a (001)-STO substrate[15], with e) and f) demonstrate the details at specific points U and R. h) displays the Fermi-surface measured from a $SrIrO_3$ film grown on a Nd-doped (001)-STO substrate[31].

|  | High-symmetry line in cubic BZ | $\frac{dE}{dk}\big|_{k=k_F}$ (eV Å) | $v_F$ ($\times 10^7$ cm/s) |
| --- | --- | --- | --- |
| SX-ARPES | X – M | 0.8 (±0.2) | 1.2 (±0.3) |
|  | M – Γ | >3.0 | >4.6 |
|  | Γ – R | 3.3 (±0.2) | 5.0 (±0.3) |
|  | R – X | 1.4 (±0.4) | 2.2 (±0.6) |
| LDA+SOC | X – M | 0.64 | 0.97 |
|  | M – Γ | 2.5 | 3.8 |
|  | Γ – R | 3.1 | 4.7 |
|  | R – X | 0.92 | 1.4 |

**Table 1** Energy dispersion gradient of the $J_{eff} = 1/2$ band at the Fermi level and the Fermi velocity in $SrIrO_3$ obtained by ARPES and theoretical calculations. Adapt from Ref. [31]

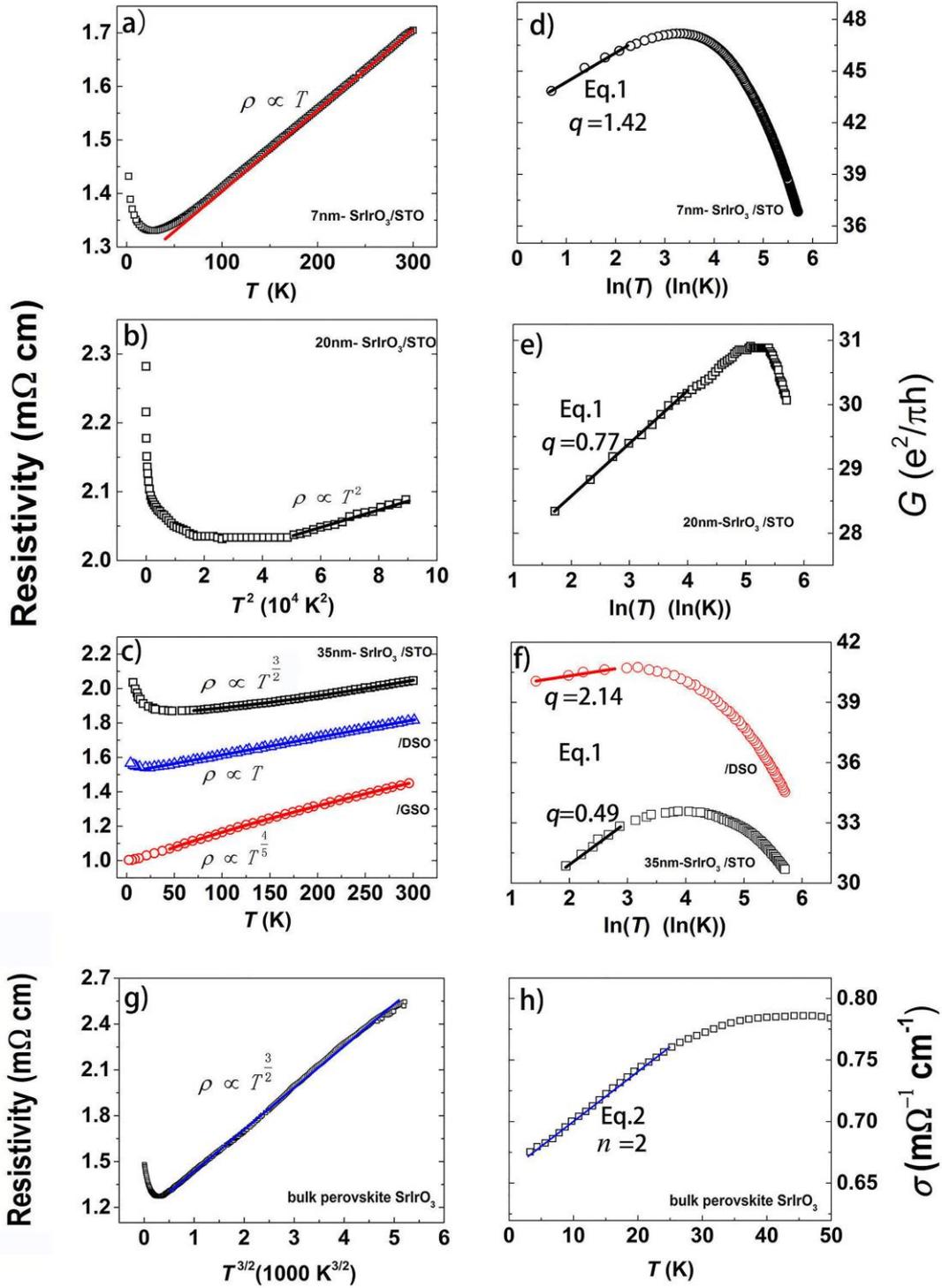

**Fig.8** Temperature-dependent resistivity of SrIrO$_3$ films with different thicknesses grown on (001)-SrTiO$_3$ substrates by various research groups [13, 17, 18, 45], showing unified features for the metal insulator transition and the non-Fermi liquid behavior.

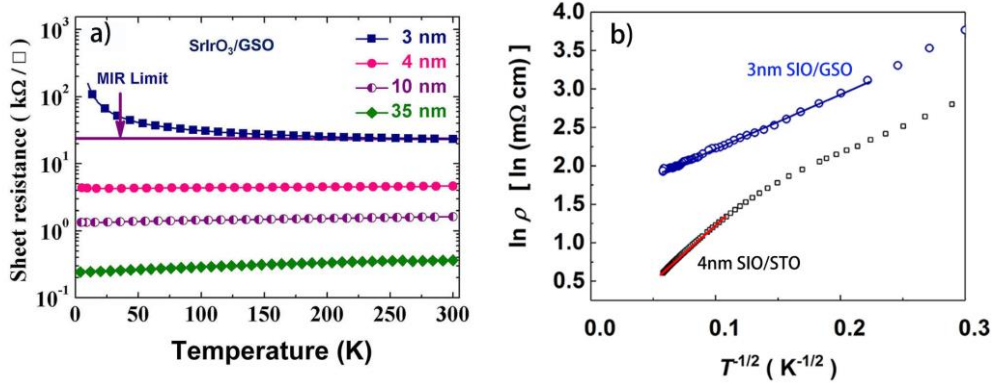

**Fig.9** Thickness dependent resistivity of SrIrO$_3$ SLFs. a) Sheet resistance of films grown on GSO substrates with varied thickness[45]; b) ln$\rho$-$T^{-1/2}$ relationship of ultrathin films at low temperature, indicating the Efros–Shklovskii VRH transport mechanism. The original data was extracted from a) for the 3nm film and from Ref. [52] for the 4nm film.

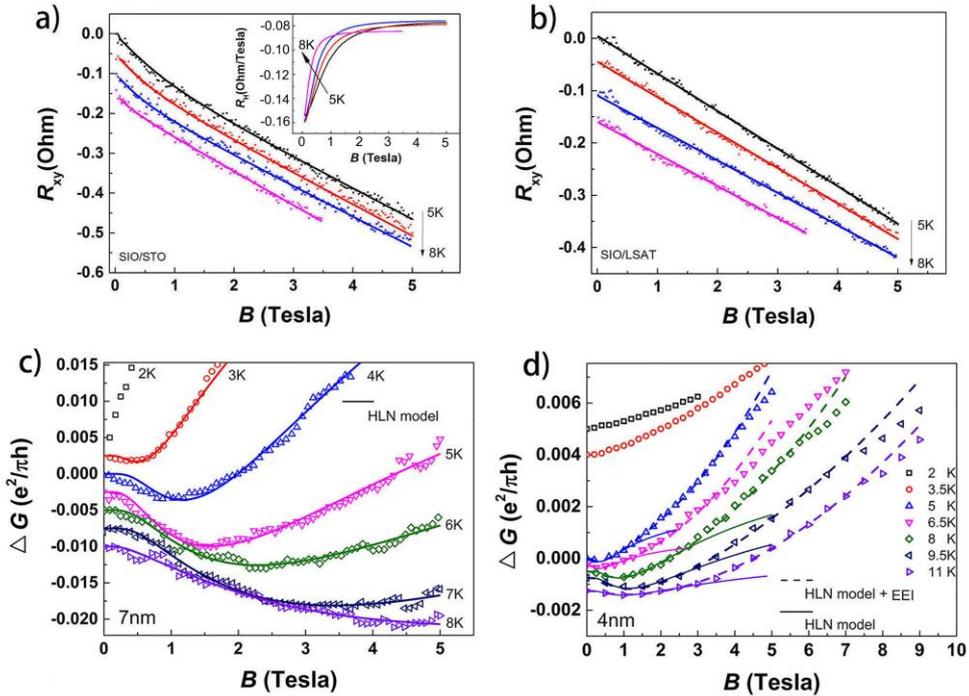

**Fig.10 a)** Nonlinear Hall effect in 7-nm thick SrIrO$_3$ films on (001)-SrTiO$_3$ and **b)** nearly linear Hall effect in the SrIrO$_3$ film on (001)-LSAT substrates [13]. The inset in a) provides the Hall coefficient of the film on SrTiO$_3$ substrate. The magnetoconductance of SrIrO$_3$ films of **c)** 7-nm and **d)** 4-nm thickness grown on (001)-SrTiO$_3$ substrates [55]. The 7-nm thick film data were fitted by the HLN model expressed by Eq.4 and the 4-nm thick film data were fitted by the HLN+EEI model shown in Eq.5.

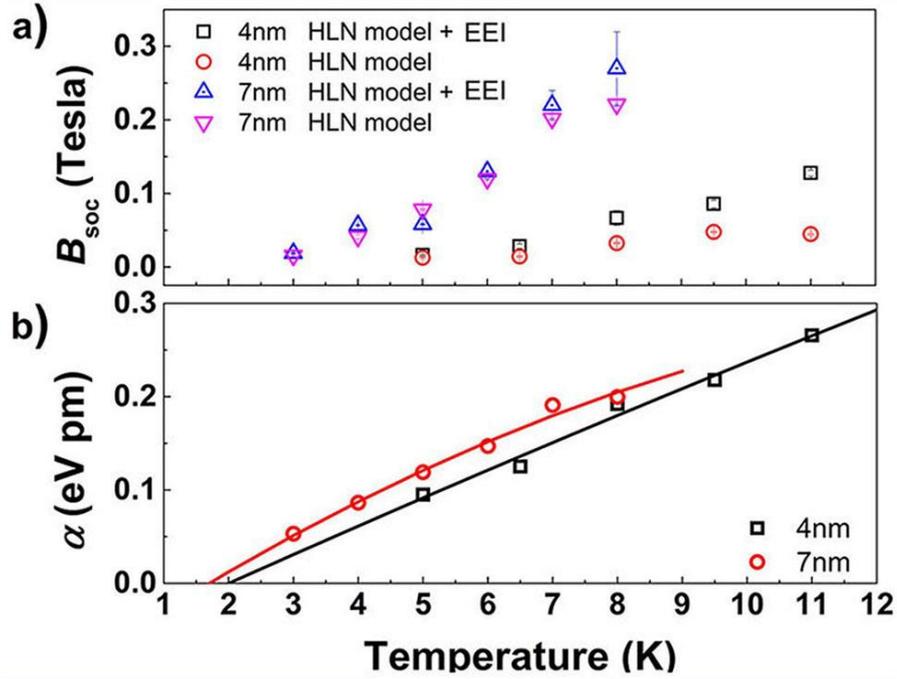

**Fig.11** Temperature dependences of **a)** the spin–orbit coupling equivalent field and **b)** the Rashba coefficient of 4-nm and 7-nm thick $SrIrO_3$ films grown on (001)-$SrTiO_3$ substrates [55]

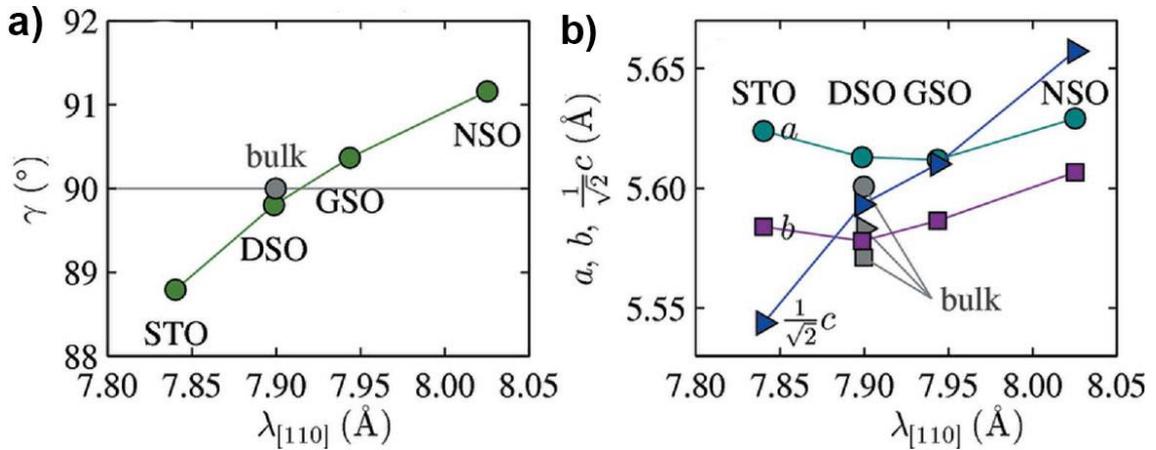

**Fig.12** Substrate strain induced structure evolution of $SrIrO_3$ films[52]. a) the $\gamma$ angle; b) lattice constants $a$, $b$ and $c$.

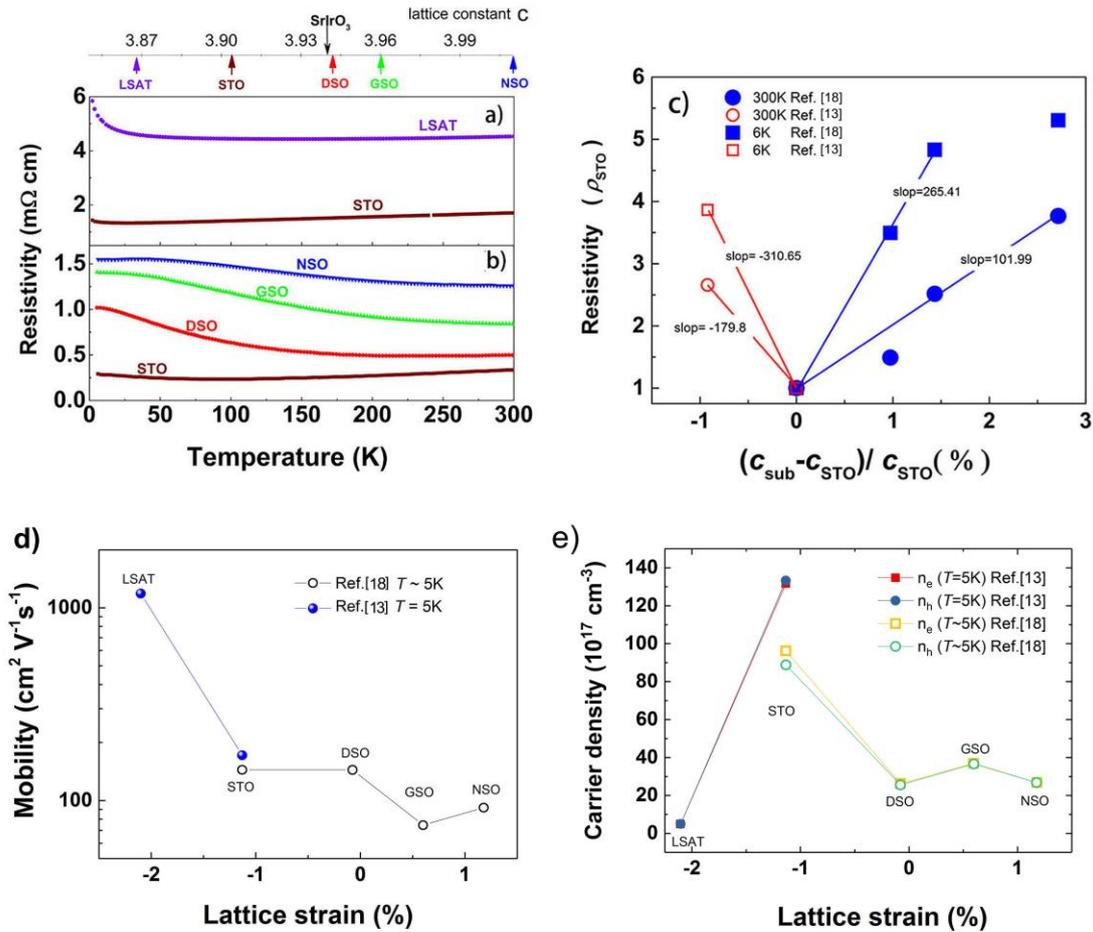

**Fig.13 a) and b)** Resistivity of $SrIrO_3$ films grown on different substrates; **c)** resistivity evolution with the substrate in plane lattice constants, where the resistivity and the lattice constant take the data for the $SrTiO_3$ substrate as references. Carrier **d)** mobility and **e)** density evolution with the in-plane lattice strain in the $SrIrO_3$ films on different substrates. (Data were reproduced from Refs. [13,18].)

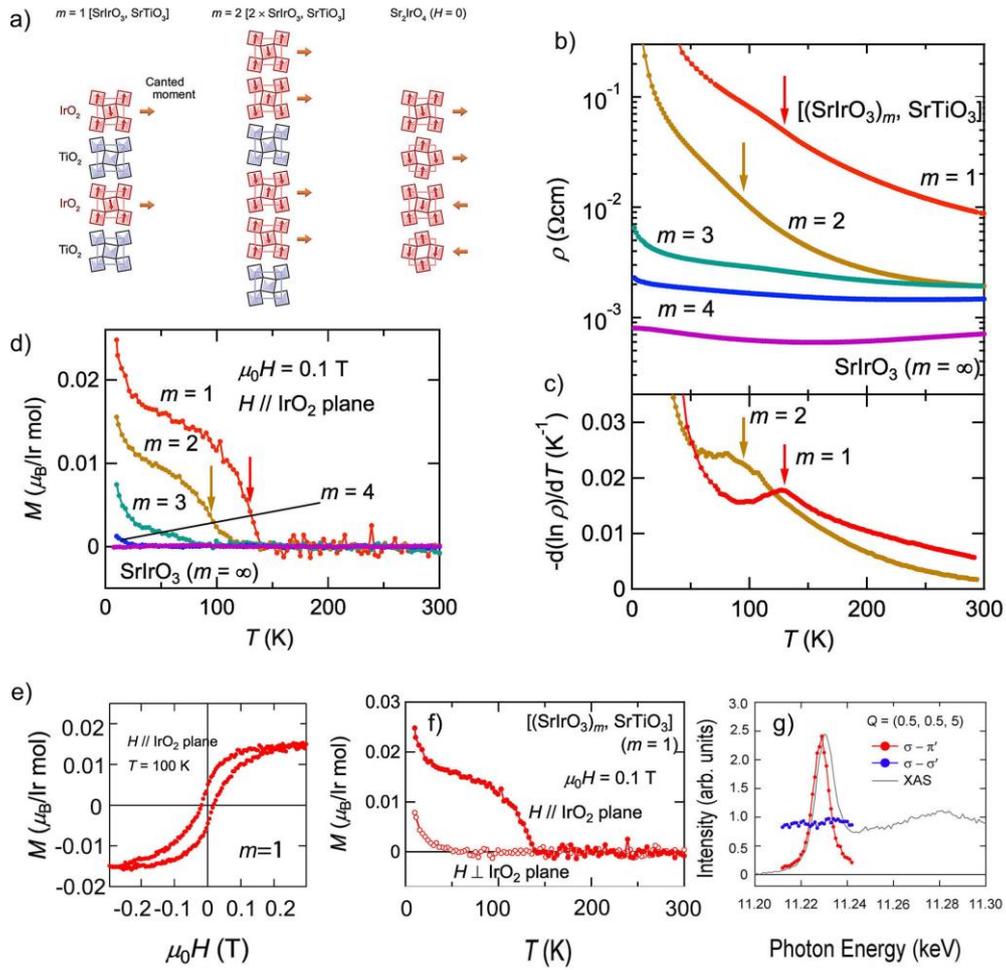

**Fig.14** Magnetic ordering transition in [(SrIrO$_3$)$_m$/SrTiO$_3$] superlattices, where $m$ represents the unit cell number of the SrIrO$_3$ layer[19]. **a)** Schematic structure of the magnetic ordering; **b) and c)** abnormal behavior of magnetic phase transition-induced resistance; **d)** Temperature-dependent magnetization and **e)** the magnetic hysteresis loop; **f)** temperature-dependent magnetization with the IrO$_2$ plane parallel and perpendicular to the magnetic fields, showing the magnetic order anisotropy; **g)** resonant magnetic X-ray diffraction of the superlattice, where only the in-plane antiferromagnetic ordering of the corresponding σ-π' polarization peak was observed.